\documentclass[twocolumn,tighten]{aastex62}

\usepackage{longtable}
\usepackage{afterpage}

\usepackage{graphicx}
\usepackage{xcolor}

\usepackage{float}
\usepackage{color}
\usepackage{amsmath}

\def\kms{$\mathrm{km \,s}^{-1}$}

\begin{document}

\shorttitle{On the Magnetic Nature of an Exploding Granule}
\shortauthors{Guglielmino et al.}

\title{On the Magnetic Nature of an Exploding Granule as Revealed by \textsc{Sunrise}/IMaX}

\author{Salvo~L. Guglielmino}
\affiliation{Dipartimento di Fisica e Astronomia ``Ettore Majorana'' -- Sezione Astrofisica, Universit\`{a} degli Studi di Catania, Via S.~Sofia 78, I-95123 Catania, Italy}

\author{Valent\'in Mart\'inez Pillet}
\affiliation{NSO - National Solar Observatory 3665 Discovery Drive, Boulder, CO 80303, USA}

\author{Basilio Ruiz~Cobo}
\affiliation{IAC -- Instituto de Astrof\'isica de Canarias, C/ V\'ia L\'actea s/n, E-38200, La Laguna, Tenerife, Spain}
\affiliation{ULL -- Departamento de Astrof\'isica, Univ.~de La Laguna, E-38205, La Laguna, Tenerife, Spain}

\author{Luis~R. Bellot Rubio}
\affiliation{IAA - Instituto de Astrof\'isica de Andaluc\'ia (CSIC), Apdo.\ de Correos 3004, E-18080 Granada, Spain}
	
\author{Jos\'e~Carlos del~Toro Iniesta}
\affiliation{IAA - Instituto de Astrof\'isica de Andaluc\'ia (CSIC), Apdo.\ de Correos 3004, E-18080 Granada, Spain}
		
\author{Sami~K. Solanki}
\affiliation{Max-Planck-Institut f\"{u}r Sonnensystemforschung, Justus-von-Liebig-Weg 3, D-37077 G\"{o}ttingen, Germany}
\affiliation{School of Space Research, Kyung Hee University, Yongin, Gyeonggi-Do, 446-701, Republic of Korea}		

\author{Tino~L. Riethm\"{u}ller}
\affiliation{Max-Planck-Institut f\"{u}r Sonnensystemforschung, Justus-von-Liebig-Weg 3, D-37077 G\"{o}ttingen, Germany}

\author{Francesca Zuccarello}
\affiliation{Dipartimento di Fisica e Astronomia ``Ettore Majorana'' -- Sezione Astrofisica, Universit\`{a} degli Studi di Catania, Via S.~Sofia 78, I-95123 Catania, Italy}

\correspondingauthor{Salvo L. Guglielmino}
\email{salvatore.guglielmino@inaf.it}

\begin{abstract}

We study the photospheric evolution of an exploding granule observed in the quiet Sun at high spatial ($\sim$ 0\farcs3) and temporal (31.5 s) resolution by the imaging magnetograph \textsc{Sunrise}/IMaX in June 2009.
These observations show that the exploding granule is cospatial to a magnetic flux emergence event occurring at mesogranular scale (up to $\sim 12 \,\mathrm{Mm}^{2}$ area). 
Using a modified version of the SIR code for inverting the IMaX spectropolarimetric measurements, we obtain information about the magnetic configuration of this photospheric feature. 
In particular, we find evidence of highly inclined emerging fields in the structure, 
carrying a magnetic flux content up to $\sim 4 \times 10^{18} \,\mathrm{Mx}$.
The balance between gas and magnetic pressure in the region of flux emergence, compared with a very quiet region of the Sun, indicates that the additional pressure carried by the emerging flux increases by about 5\% the total pressure and appears to allow the granulation to be modified, as predicted by numerical simulations. 
The overall characteristics suggest that a multi-polar structure emerges into the photosphere, resembling an almost horizontal flux sheet. This seems to be associated with exploding granules. Finally, we discuss the origin of such flux emergence events.

\end{abstract}


\section{Introduction}

A typical phenomenon observed in the quiet-Sun granulation pattern is the appearance of \textit{exploding granules} (EGs). They are individual bright granules that expand more than normal granular cells, and ultimately fragment into several smaller granules. They have a rather long lifetime, with an average value of $\approx 9$ minutes, and reach a maximum diameter of about $4\arcsec-5\arcsec$ \citep{Mehl:78,Title:89,Rast:95}. 

Since the first detection of EGs \citep{Rosch:60}, a number of studies analyzed their dynamics and physical properties, from both the observational point of view \citep{Carlier:68,Namba:69,Allen:73,Mehl:78,Namba:77,Kitai:79,Namba:86,Title:89,Hirz:97,Hirz:99a,Hirz:99b,Roudier:01,Hirz:01,Berrilli:02,Santiago:10,Palacios:12,Sobotka:12,Fischer} and numerical modeling (e.g., \citealp{Musman:72,Nelson:78,Nordlund:85,SimonWeiss:91,Rast:95,Stein:98}; see also the review of \citealp*{Nordlund:09} and the recent simulations by \citealp{Moreno:18,Rempel:18}).

Regular granules can be divided into two populations, with diameters smaller than and larger than 1\farcs4, respectively \citep[e.g.,][]{Hirz:97,Gadun:00,Berrilli:02,Yu:11}. Small and large granules exhibit differences in the geometrical and brightness structure and in the temporal evolution \citep{Hirz:97,Hirz:99a}. EGs belong to the family of large granules, as they die by fragmentation and have the longest mean lifetime. Like in large granules, also in EGs the brightest parts and maximum upward velocity are shifted towards the granular boundaries \citep{Hirz:02}. Temporal sequences of white-light, high-resolution observations showed that EGs are not rare, as their number density is about 4\% of the observed area of the photosphere \citep{Mehl:78}. Later works found slightly lower values: EGs covering about the 2.5\% of the photosphere \citep{Namba:86,Title:89}. 

EGs often exhibit a central dark spot, as first described by \citet{Kitai:79}. The possible connection of this dark hub with downflows of cool gas within the granules was suggested by \citet{Hirz:99b}. Two-dimensional spectroscopic measurements confirmed the existence of such downflows in the central area \citep{Hirz:01,Roudier:01,Berrilli:02}. The dark center of EGs is interpreted to be a result of buoyancy braking \citep{Massaguer:80}. The pressure near the center of granules increases while sustaining their horizontal radial flows to conserve mass. The enhanced pressure reduces the upflow and heat transport to the surface. This process produces a stagnation point, which rapidly cools down and eventually reverses the flow, leading to the formation of the observed dark central feature, which is characterized by downward velocity while the EG starts splitting \citep{Nelson:78}.
 
The explosion of an EG can occur in a recurrent way, with some fragments that explode again \citep{Oda:84,Namba:86}. For that reason, EGs represent the most vigorous manifestation of fragmenting granules \citep{Roudier:03}. This behaviour was already noticed by \citet{Kawaguchi:80} for granules with diameters greater than $2\arcsec$, but for EGs it was further related to mesogranulation in observations \citep[e.g.,][]{Rieutord:00,Berrilli:05} and numerical models \citep{November:81,Oda:84,Simon:91,Ploner:00}. In particular, it was observed that small convective elements (granules) superimposed on larger cells (mesogranules), randomly distributed, may give rise to the formation of EGs \citep{SimonWeiss:91}. In this context, areas of positive divergence of the horizontal velocity, 
which form the mesogranular pattern found in hydrodynamical simulations \citep{Rast:03,Matloch:09,Matloch:10}, are identified with EGs in observations \citep[e.g.,][]{Domi:03,Roudier:03,Roudier:04}.

In spite of the detailed knowledge of the morphological evolution of EGs, their correlation with magnetic flux emergence episodes is not well settled. \citet{DePontieu:02} found the presence of emerging flux in certain EGs, suggesting the presence of mostly horizontal magnetic fields. He related these flux emergence events with the so-called horizontal internetwork fields (HIFs) discovered by \citet{Lites:96}. Also \citet{Hector:04}, analyzing low-flux signals in internetwork regions, pointed out a scenario compatible with emergence of flux at granular scale. More recently, \citet{Orozco:08} showed individual cases of magnetic flux emergence in granules with full Stokes polarimetry at high spatial and temporal resolution. \citet{Zhang:09} observed flux emergence as a cluster of mixed polarities following
the splitting of a large granule. \citet{Palacios:12} investigated the evolution of two mesogranular-scale EGs, finding weak unipolar longitudinal fields appearing first. These magnetic flux concentrations were followed by the appearance of the opposite polarity and developed into intergranular lanes, while the transverse field remained almost negligible in the flux concentration where it could be measured.

In this paper, we benefit from observations of the solar photosphere taken by the Imaging Magnetograph eXperiment \citep[IMaX;][]{Valentin:11} on board the \textsc{Sunrise} balloon-borne solar observatory \citep{Solanki:10,Solanki:17,Barthol:11,Berkefeld:11,Gandorfer:11}. 
The high spatial resolution and polarimetric sensitivity of the magnetograms acquired by IMaX shed light on the evolution of several kinds of small-scale magnetic features in the quiet Sun. Indeed, it provided a rich picture of the dynamic processes that involve individual magnetic elements in the photosphere (e.g., emergence of small loops, \citealp{Danilovic:10,Guglielmino:12}; convective collapse, \citealp{Requerey:14,Requerey:15}; dynamics of magnetic bright points, \citealp{Jafarzadeh:13,Jafarzadeh:14,Utz:14}; for a statistical analysis, see \citealp{Anusha:17}). Furthermore, IMaX measurements allowed magnetic flux tubes to be spatially resolved in the internetwork \citep{Lagg:10} and network \citep{Marian:12}. 

In the present work, we analyze at high spatial and temporal resolution the evolution of an EG, which is cospatial to a flux emergence event occurring at mesogranular scale. We study this emerging flux region, which was already described by \citet{Palacios:12}, complementing the analysis with new information concerning the characterization and distribution of the polarization signals across the feature and its thermodynamical properties. 
Sect.~2 describes the observations and the methods that we used for the data analysis. In Sect.~3 we report our results, discussing them in comparison with magnetohydrodynamical (MHD) simulations in Sect.~4. Finally, Sect.~5 summarizes our results, putting them in a more general context.

\section{Observations, Methods,\\and Data analysis}

\begin{figure}[t]
	\centering
	\includegraphics[trim=25 45 30 30, clip, scale=0.45]{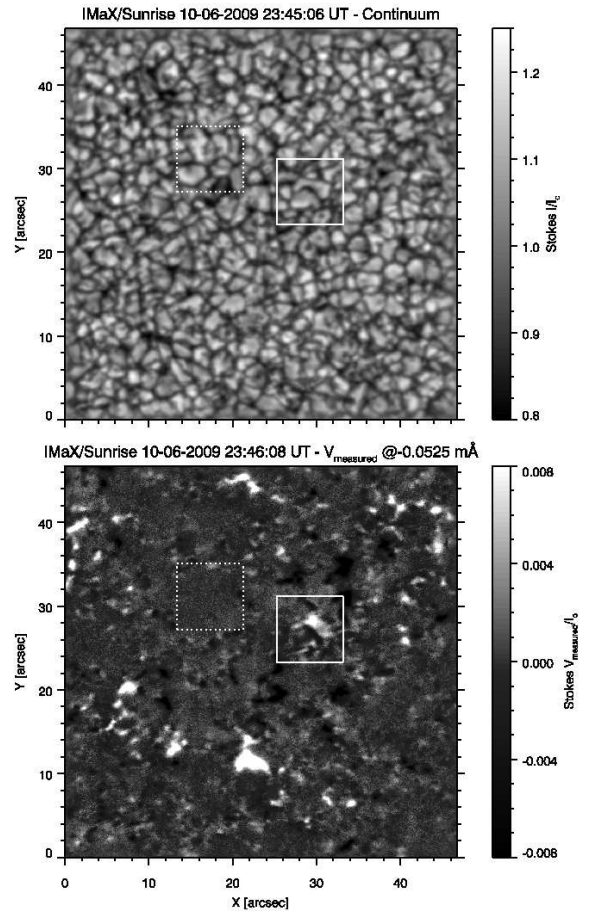}
	\caption{\emph{Top panel}: Continuum intensity map. \emph{Bottom panel}: $V_{\mathrm{measured}}$ signal map in the blue lobe of the observed Stokes~\emph{V}. They cover the FoV of IMaX of about $46\arcsec \times 46 \arcsec$. 
	The solid square, with a subFoV of $\sim 8\arcsec \times 8\arcsec$, indicates the location of the emerging flux region. The dashed square, with the same subFoV, indicates a region of very quiet Sun, used as a control box for comparison. \label{fig0}}
\end{figure}

We have analyzed a spectropolarimetric IMaX data set obtained on 2009 June 10, 22:55:41--23:54:30~UT, during the first science flight of the \textsc{Sunrise} observatory. During that time, IMaX took polarization maps at twelve wavelength positions over the \ion{Fe}{1} 525.02~nm line (Land\'e factor $g=3$) every 3.5~pm from $\lambda = -19.25$~pm to $\lambda = +19.25$~pm with respect to the line center, with a spectral resolution of 8.5~pm. Two images were accumulated per wavelength point, performing measurements of the Stokes parameters~\emph{I} and~\emph{V} only (longitudinal L12-2 observing mode). The pixel size of the maps is 0\farcs055. The field of view (FoV) covered by these observations is about $46\arcsec \times 46\arcsec$ over a quiet Sun region at the disc center, as shown in Figure~\ref{fig0}. The temporal cadence of acquisition was 31.5~s. 

Data have been processed for dark-current subtraction and flat-field correction. Two different types of data have been produced: non-reconstructed (level~1) and reconstructed (level~2), the latter being obtained by using phase-diversity information and taking into account the wave-front correction \citep{Berkefeld:11}. The spatial resolution is 0\farcs3 (level~1) and 0\farcs23 (level~2), respectively. The same data set was analyzed by \citet{Tino:14} to study bright points. We refer the reader to that paper for information on the data and the observed region. Further details about data reduction are provided by \citet{Valentin:11}. In the following we use the level-1 data, which ensures a signal-to-noise ratio of about $1.2 \times 10^{-3}$ in units of the continuum intensity $I_c$ per wavelength point in Stokes~\emph{V}. The rms contrast of the granulation in the continuum is 7\% for these level-1 data.

$I_c$ has been computed by averaging the Stokes~\emph{I} intensities at $\lambda = \pm 19.25$~pm from the line center. The resulting $I_c$ maps have been used to align the observational sequence referring to the flux emergence event. Actually, some slight displacements of the FoV occurred during the observing interval, due to short glitches of the image stabilization system. Therefore, we have used a cross-correlation algorithm to track the EG between the observational gaps. The total length of the analyzed time series is about 17.5~min.

\subsection{Polarization cross-talk}

While using longitudinal IMaX observing modes, such as the L12-2 mode, it is not possible to correct for instrumental cross-talk by applying the demodulation matrix described by \citet{Valentin:11}, which takes into account the polarization effect induced by the telescope and by the Image Stabilization and Light Distribution system \citep[ISLiD;][]{Gandorfer:11}. This implies that a conspicuous cross-talk between the Stokes parameters still remains in the reduced data. 

An estimate of the cross-talk for Stokes~\emph{V}, obtained from a measurement performed just before the flight of the \textsc{Sunrise} balloon, is given by 

\begin{equation}
V_{\mathrm{measured}} = - 0.88 \, U_{\mathrm{source}} + 0.55 \, V_{\mathrm{source}} \;,\label{eqUVsource}
\end{equation}


\noindent
where $U_{\mathrm{source}}$ and $V_{\mathrm{source}}$ refer to the calibration source. The cross-talk originates mostly from the two folding mirrors behind the primary mirror, M3 and M4, and the ISLiD. Hence, the \textit{observed} Stokes~\emph{V} signal does not represent the real circular polarization signal. 

An additional problem is that we lack pointing information for these IMaX observations, so we cannot set a reference frame for the Stokes vector in which Stokes~$Q_{\mathrm \sun}=0$, e.g., at solar West, indicating with subscript $\sun$ the \textit{original} solar signals in such a reference frame. That is, in general we should represent the emerging observed Stokes~\emph{Q} and~\emph{U} parameters as a linear combination of Stokes~$Q_{\mathrm \sun}$ and~$U_{\mathrm \sun}$, respectively. 
	
%
%
%
%
%


In general, there is a rotation between $Q_{\mathrm{source}}$ and $U_{\mathrm{source}}$ to $Q_{\mathrm \sun}$ and $U_{\mathrm \sun}$ that one needs to know carefully to provide magnetic field directions on the plane of the sky. However, as shown in Section~\ref{UVcode} and in the Appendix, the inversion procedure we use does not determine the azimuth (or $B_{\mathrm{trans}}$) in any reliable way. Thus, the additional rotation to provide orientations in the plane of the sky is not used in this paper, and without losing generality, we can, therefore, write Equation~(\ref{eqUVsource}) as

\begin{equation}
V_{\mathrm{measured}} = - 0.88 \, U_{\mathrm \sun} + 0.55 \, V_{\mathrm \sun} \;.\label{eqUV}
\end{equation}

Equation~(\ref{eqUV}) will describe the \textit{observed} Stokes~\emph{V} signal as a linear combination of Stokes~$U_{\mathrm \sun}$ and~$V_{\mathrm \sun}$ that is suitable for investigating linear and longitudinal polarization signals in the EG. 
	
Typically, in the quiet Sun, the Zeeman effect produces only weak linear polarization signals. As a consequence, to zero-th order the contribution of Stokes~$U_{\mathrm \sun}$ can be neglected and $V_{\mathrm{measured}}$ can be considered as a measure of the original Stokes~$V_{\mathrm \sun}$, with a reduced amplitude by a factor of about 1.8. This assumption is a reasonable guess, for instance in network elements where vertical fields are expected, as illustrated by \citet{Marian:12}. A clear relationship between inclination and field strength is shown by \citet{Tino:17}: only weak fields are horizontal and their Stokes~$Q$ and~$U$ signals are really small (see also \citealt{Danilovic:10} for the typical strength of Stokes~$Q$ and~$U$ signals). Conversely, this assumption is broken in regions close to neutral lines, where Stokes~$U_{\mathrm \sun}$ becomes significant compared to Stokes~$V_{\mathrm \sun}$.   

Let us now consider the integrated signal of $V_{\mathrm{measured}}$ averaged over the line, given by

\begin{equation}
V_{\mathrm{integrated}}=\frac{1}{8\,\left\langle I_c\right\rangle} \sum_{i=1}^{8} \epsilon_{i} V_{\mathrm{measured}} {\big|_i} \;,\label{eqVint}
\end{equation}

\noindent
where $\left\langle I_c\right\rangle$ is the continuum intensity averaged over the IMaX FoV, $i$ runs over the central eight wavelength positions, from $\lambda$ = -12.25 pm to $\lambda$ = +12.25 pm with respect to the line center, with $\epsilon=1$ for the first 4 positions and $\epsilon=-1$ for the other 4 positions.  
As a first approximation, this quantity is null only in the regions where the magnetic field is zero or is horizontal (changing sign), i.e., where the integrated Stokes~$V_{\mathrm \sun}$ signal is zero, as long as Stokes~$U_{\mathrm \sun}$ is symmetric in wavelength while Stokes~$V_{\mathrm \sun}$ antisymmetric \citep{Egidio}. In this respect,  $V_{\mathrm{integrated}}$ represents fairly well the distribution of the magnetic areas. Neutral lines derived with this method, although being regions that are more affected by the cross-talk, are a good proxy for the real neutral lines. However, we cannot have any knowledge of the magnetic flux from Equation~(\ref{eqVint}).



\subsection{SirUV code}
\label{UVcode}

The presence of a residual cross-talk for Stokes~\emph{V} prevents us from using the standard procedures of data analysis. In fact, the inversion of the spectra as they are measured would fail to retrieve the correct values of various physical parameters (i.e., magnetic flux, magnetic field strength $B$, inclination $\gamma$\footnote{The inclination is measured relative to the line of sight, so that $0^{\circ}$ indicates a magnetic field vector pointing to the observer, i.e., directed outward from the solar surface at the disc center.}).

In order to extract more information from the IMaX spectropolarimetric measurements, we have carried out a non-standard inversion of the observed level-1 spectra by using a modified version of the SIR code \citep{RuizIniesta:92}. This code provides a numerical solution to the radiative transfer equation along the line of sight (LOS) for Zeeman-polarized radiation under the assumption of local thermodynamic equilibrium, minimizing the differences between the observed and the computed synthetic Stokes profiles using response functions. The present version, hereafter \emph{SirUV}, allows the user to invert any linear combination of the four Stokes parameters.

In the Appendix, we have assessed the capabilities of this \emph{SirUV} code using IMaX V5-6 (full vector) observations. In summary, this analysis shows that the code retrieves reliable values for the thermodynamical parameters, like the temperature $T$ and the gas pressure $P_{\mathrm{gas}}$. In addition, the longitudinal component of the magnetic field $B_{\mathrm{long}}$ is reasonably well determined. Conversely, $B_{\mathrm{trans}}$ remains completely undetermined and $B$ is not determined to any useful level, with a tendency for it to be overestimated, likely because of the noise \citep{Borrero:11,Borrero:12}.

\subsection{Data Analysis}

After verifying the capabilities of the \emph{SirUV} code and the reliability of the results that are inferred, as explained in the Appendix, we have carried out the inversion of two subFoVs of $144 \times 144$ pixels. The first (FoV$_{EG}$) covers the EG, the second (FoV$_{QS}$) is occupied by a region of very quiet Sun, which can be termed as a dead-calm region \citep{Marian:12_dead}, as indicated in Figures~\ref{fig0} with solid and dashed boxes, respectively. All the frames relevant to FoV$_{EG}$ have been inverted, using the maximum cadence of 31.5 s, while only every fifth frame relevant to FoV$_{QS}$ has been inverted, with a cadence of $\sim 2.5$ minutes, during the observing period when it was included in the full IMaX FoV. 

As regards the stratification and the number of nodes used for the inversion, we have taken the same parameters used in the preliminary test. That is, we have used 3 iteration cycles, with up to 4 nodes for the temperature $T$, 2 nodes for the LOS velocity $v_{\mathrm LOS}$ and $B$, and 1 node for the other parameters. 
Under the assumption of unity for the magnetic filling factor, in the following we consider the values of $B_{\mathrm{long}}$ in terms of magnetic flux density. 

Given the large uncertainty of the $v_{\mathrm{LOS}}$ values retrieved by the \emph{SirUV} inversion, the LOS velocity has been derived from a Gaussian fit to the twelve points along the observed Stokes~\emph{I} profile, considering the Doppler shift with respect to the line center. The blueshift over the FoV due to the collimated setup of the Fabry-P\'erot etalon of the magnetograph has been also removed in the inferred velocity values. Taking into account that the random superposition of solar acoustic oscillation coherence patches and underlying convective flows may introduce a bias in the interpretation of the Doppler velocities in IMaX data \citep[see][]{McClure:19}, we have applied a \emph{p}-mode filtering. Finally, we have calibrated the velocity scale over the entire FoV by using a convective blueshift \citep{Dravins:81} of $\sim 200\;\textrm{m s}^{-1}$ for the \ion{Fe}{1} 525.02 nm line at disc center, as used for other IMaX data sets \citep[see, e.g.,][]{Roth:10}. The typical error of this measure is about $\pm 100 \,\mathrm{m s}^{-1}$. 

We have also studied the behaviour of $T$, $P_{\mathrm{gas}}$, and of the magnetic pressure $P_{\mathrm {mag}} = {B^{2}}/{2\mu_0}$, where $\mu_0$ is the vacuum permeability. To estimate $P_{\mathrm {mag}}$, we have considered that in the emerging flux region it can be assumed that

\[
B_\mathrm{tran} \simeq B_\mathrm{long}
\]

\noindent
thus

\[
P_{\mathrm{mag}} = \frac{B^{2}}{2\mu_0} = \frac{B_\mathrm{long}^{2} + B_\mathrm{tran}^{2}}{2\mu_0} \simeq \frac{B_\mathrm{long}^{2}}{\mu_0} \; .
\]

\noindent
Hence, the quantity depends only on $B_\mathrm{long}$, which is a well determined observable from our spectropolarimetric measurements. 

The spatial averages of $T$, $P_{\mathrm{gas}}$, and $P_{\mathrm {mag}}$ have been carried out separately for granules and intergranular lanes. For this purpose, we applied a discrimination based on the continuum brightness of these features. Pixels brighter than their surroundings, considering the average value of a $19 \times 19$ pixels box surrounding the point under consideration, by at least 4\%, are taken to be granules. In the same way, the criterion is used to identify intergranular lanes, as pixels that are at least 3\% darker than their surroundings.

\section{Results}

\begin{figure*}[t]
	\centering
	\includegraphics[trim=10 40 0 5, clip, scale=0.795]{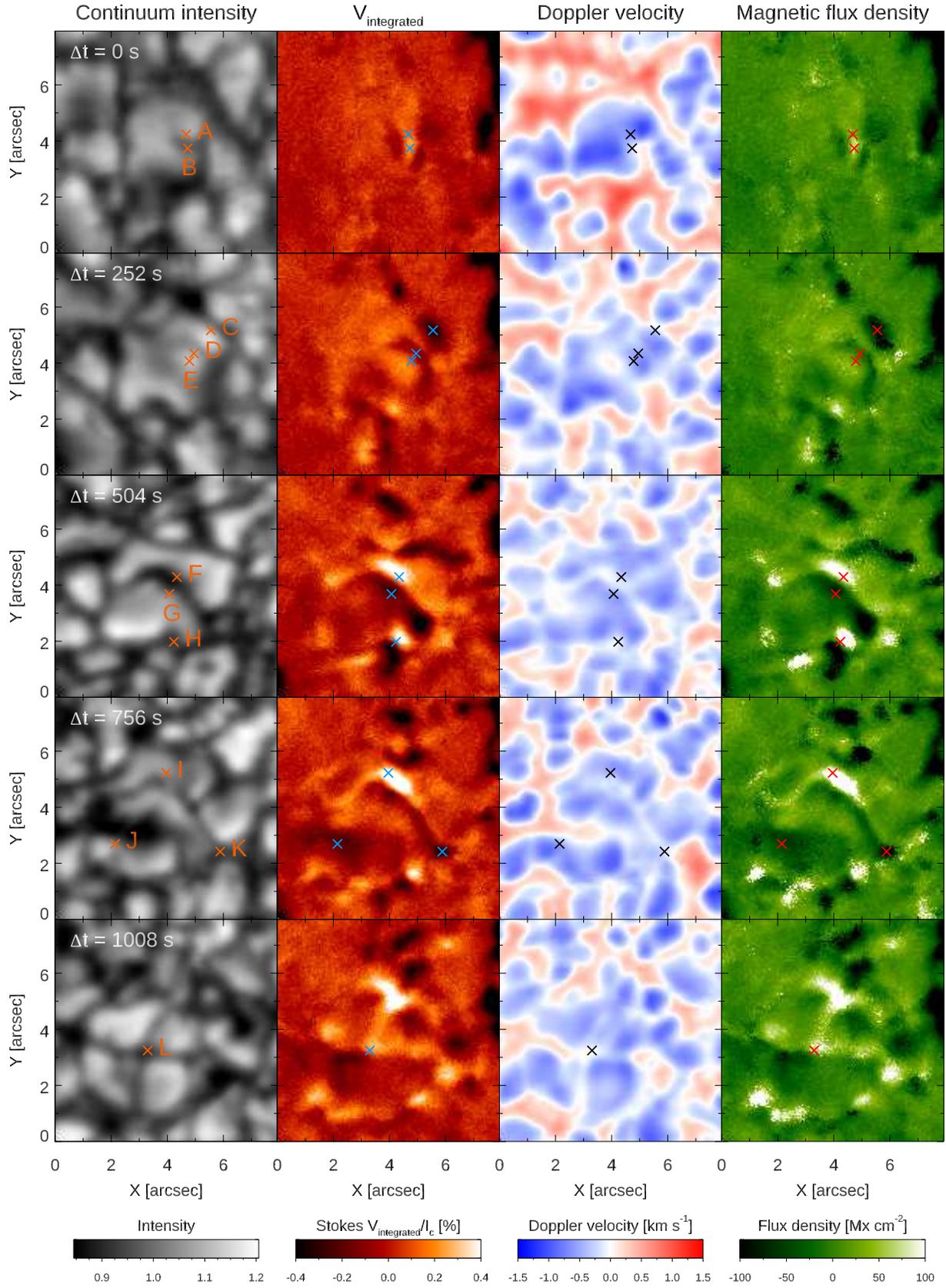}
	\caption{Time sequence of the evolution of the emerging magnetic structure. The FoV is that framed with a solid line box in Figure~\ref{fig0}. \textit{First column}: Continuum intensity maps. \textit{Second column}: Observed circular polarization ($V_{\mathrm{observed}}$) maps. \textit{Third column}: Doppler velocity $v_{Dopp}$ maps. \textit{Fourth column}: Magnetic flux density maps. Time runs from top to bottom: the cadence of the sequence is about 4 minutes. Crosses labeled with capital letters indicate pixels whose Stokes profiles are shown in Figure~\ref{fig6}. An animation of this figure is available in the online journal. \label{fig4}}
\end{figure*}

A magnetic flux concentration with a complex mixed polarity pattern is seen to emerge at 23:36:42 UT ($\Delta t = 0$~s). The feature is clearly recognizable in the full FoV of these IMaX L12-2 observations, as shown in Figure~\ref{fig0} about nine minutes later. This flux emergence episode, described as ``emergence event \emph{l10}'' in \citet{Palacios:12}, is found to be associated with an EG. 

Figure~\ref{fig4}, covering the FoV$_{EG}$, illustrates the evolution of the EG and of the cospatial emerging flux region, with a cadence of about 4 minutes beginning from $\Delta t = 0$~s. The movie available in the online journal clearly displays the intermediate phases during the evolution of the structure, with the highest temporal cadence (31.5~s), except for two observational gaps. The first gap occurred between  $\Delta t \approx 120$~s and $\Delta t \approx 240$~s, whereas the second one, shorter than the former, lies between $\Delta t = 315$~s and $\Delta t = 378$~s.

The first column of Figure~\ref{fig4} presents the evolution of the photospheric continuum $I_c$. We can observe that this expanding structure begins to form a central dark spot ($\Delta t = 252$~s), which eventually grows in size while the EG splits into two smaller granules ($\Delta t = 504$~s). Then, the southernmost descendant granule located at $\left[2\arcsec, 4\arcsec\right] \times \left[2\arcsec, 4\arcsec\right]$ undergoes the same evolution ($\Delta t = 756$~s), until it splits into four granules ($\Delta t = 1008$~s). 

The second column of Figure~\ref{fig4} shows the integrated signal of $V_{\mathrm{measured}}$. The emerging flux region first appears at $\Delta t = 0$~s, being located at [5\arcsec, 4\arcsec]. Magnetic field patches with apparent opposite polarity appear close to each other. These magnetic patches expand and occupy the EG area ($\Delta t = 252$~s), showing a pattern of alternate polarities (\textit{salt-pepper} or \textit{sea-serpent} configuration). Some of these flux kernels intensify: notably, this occurs in the region where the first dark spot is forming ($\Delta t = 504$~s). Later, this structure becomes an intergranular lane ($\Delta t = 756$~s).

Doppler velocity maps, displayed in Figure~\ref{fig4} (third column), highlight the plasma motions in the region. They indicate that the structure is generally characterized by conspicuous blueshifts, corresponding to upflows up to $1-1.5$~\kms. In the central dark patch downflows of $\simeq 0.5$~\kms{} are found ($\Delta t = 504$~s and, partly, at $\Delta t = 1008$~s), in agreement with previous observations. The same coherent behaviour is visible in the online movie as well.

The maps of magnetic flux density (Figure~\ref{fig4}, fourth column) reveal that the distribution of the magnetic areas derived from $V_{\mathrm{integrated}}$ is representative of the real magnetic flux patches. Indeed, a mixture of polarities is found where $V_{\mathrm{integrated}}$ maps show an apparent \textit{salt-pepper} configuration, suggesting once again that the quantity described in Eq.~\ref{eqVint} is a good qualitative proxy for the longitudinal component of the magnetic field, $B\,\cos\gamma$. The evolution of magnetic flux density qualitatively follows that of $V_{\mathrm{integrated}}$, as mentioned above.

\begin{figure}[t]
	\centering
	\includegraphics[trim=45 105 150 60, clip, scale=0.525]{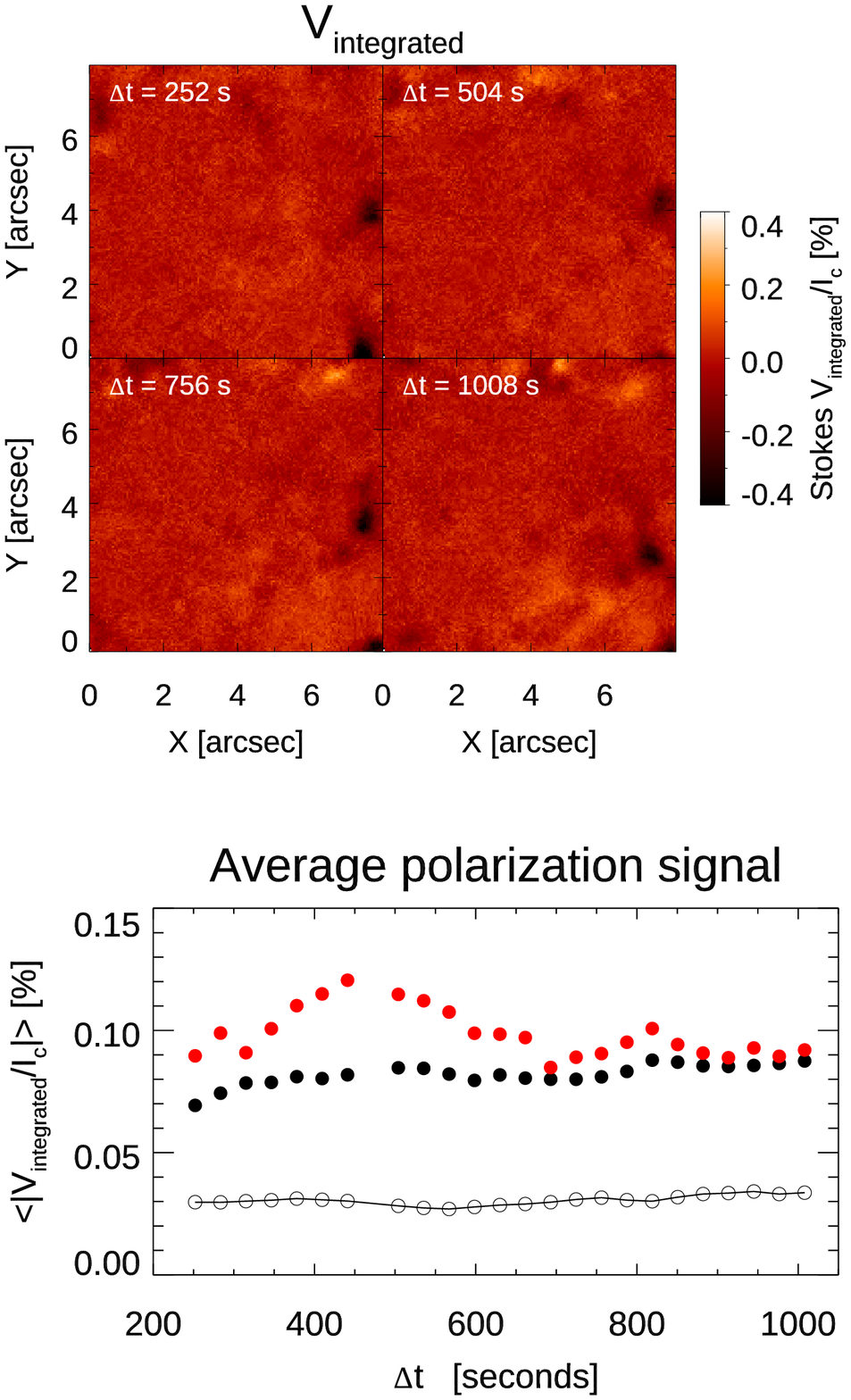}
	\caption{\emph{Top panels}: Observed circular polarization ($V_{\mathrm{observed}}$) maps at representative times for FoV$_{QS}$, that is framed with a dashed line box in Figure~\ref{fig0}. \emph{Bottom panel}: Mean value of absolute $V_{\mathrm{integrated}}$ over FoV$_{EG}$ (black circles), within the region of interest occupied by the EG inside FoV$_{EG}$ (red circles), and FoV$_{QS}$ (empty circles). \label{figQS}}
\end{figure}

\begin{figure}[!b]
	\centering
	\includegraphics[trim= 20 550 280 85, clip, scale=.57]{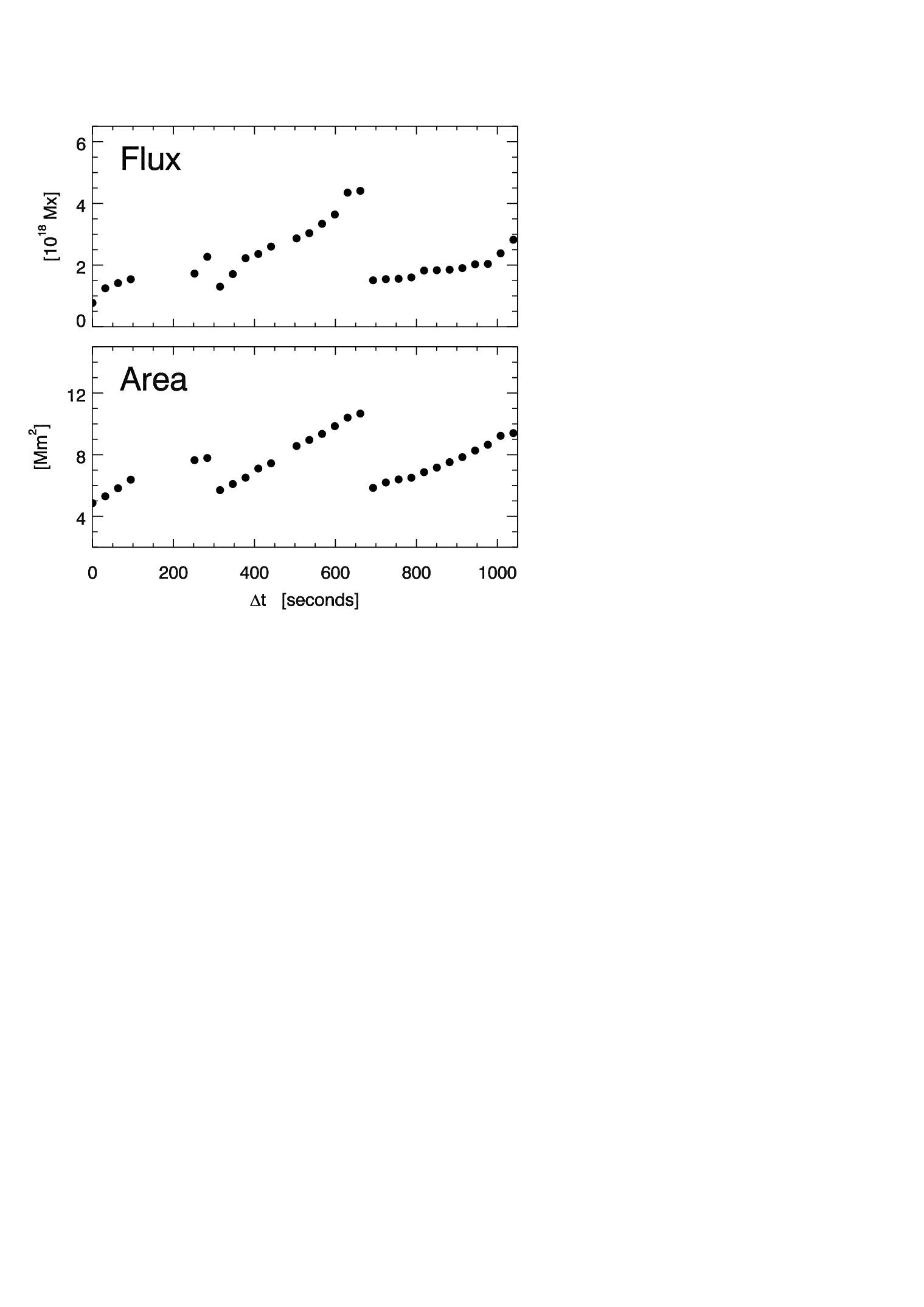}
	\caption{\textit{Top panel}: Plot of the evolution of the total (unsigned) magnetic flux content in the emerging structure cospatial to the EG in the FoV$_{EG}$. \textit{Bottom panel}: Same, for the area covered by magnetic flux emergence. \label{fig5}}
\end{figure}

A polarimetric analysis of the dead-calm region \citep[see][]{Marian:12_dead} in FoV$_{QS}$, used as a control box for comparison, is presented in Figure~\ref{figQS}. This region of very quiet Sun entered the IMaX FoV after the first 200~s of the EG evolution. In Figure~\ref{figQS} (top panels) we present the maps of the integrated signal of $V_{\mathrm{measured}}$, at the same representative times shown for FoV$_{EG}$ in Figure~\ref{fig4} (second column). No new magnetic structures are seen to emerge during the analyzed time series. Figure~\ref{figQS} (bottom panel) shows a plot of the mean polarization signal $|V_{\mathrm{integrated}}|$ averaged over FoV$_{EG}$ (black symbols), within the region of interest occupied by the EG inside FoV$_{EG}$ (red symbols), and over FoV$_{QS}$ (empty circles), along the sequence. The signal in FoV$_{EG}$ is on average 2.7 times stronger than in FoV$_{QS}$. Moreover, the polarization signal in FoV$_{QS}$ remains almost constant during the time series, while in FoV$_{EG}$ it shows an enhancement, even more evident in the region of interest occupied by the EG, which is correlated with the magnetic flux increase. This demonstrates that the field emergence in the EG is significant.

Figure~\ref{fig5} displays the evolution of two parameters that describe the EG: the magnetic flux content (top) and area (bottom). To obtain these quantities, for each frame we have defined the region of interest occupied by the EG within the IMaX FoV$_{EG}$ by hand-drawn contours. 
There is, of course, a certain degree of subjectivity in defining such a region of interest, because of i) the choice of the instant at which the pre-existing structure can be definitely considered split into two or more descendant granules, and of ii) the possible inclusion of some adjacent pixels of the internetwork, which do not belong to the structure. Nonetheless, the subjectivity in our procedure does not significantly alter the results of the present analysis, as found when repeating the analysis including some additional pixels, or removing some of them. Arrays containing the indices of the pixels within the hand-drawn contours have been stored to allow future replication of the analysis.

The total (unsigned) magnetic flux content of the emerging structure grows up to $4 \times 10^{18}$~Mx, where $\mathrm{Mx} = 10^{-8}~\mathrm{Wb}$ is the CGS unit of magnetic flux, at its maximum expansion (see Figure~\ref{fig5}, top panel). From the slope of the curve, we find a steeper increase during the first 400~s ($1.5 \times 10^{18}$~Mx), then a more gradual trend, and finally a smoother increase that begins at 700~s and lasts until the end of the observations. Note that our assumption of unity filling factor has no impact on the estimate of the magnetic flux density and on the total flux content. 
In Figure~\ref{fig5} (bottom panel) we plot the area over which magnetic flux emerges. The EG covers an area of $8 - 10 \,\mathrm{Mm}^{2}$, with a monotonic increase that is almost linear in time, reaching a maximum value of about $12 \,\mathrm{Mm}^{2}$. 
The values for the diameter of the EG (obtained under the assumption that it has an ideal circular shape) fall in the range of about $4\arcsec - 5\arcsec$, comparable with those reported in the literature for EGs.

\begin{figure*}[htbp]
	\centering
	\includegraphics[trim=0 85 0 5, clip, scale=.1925]{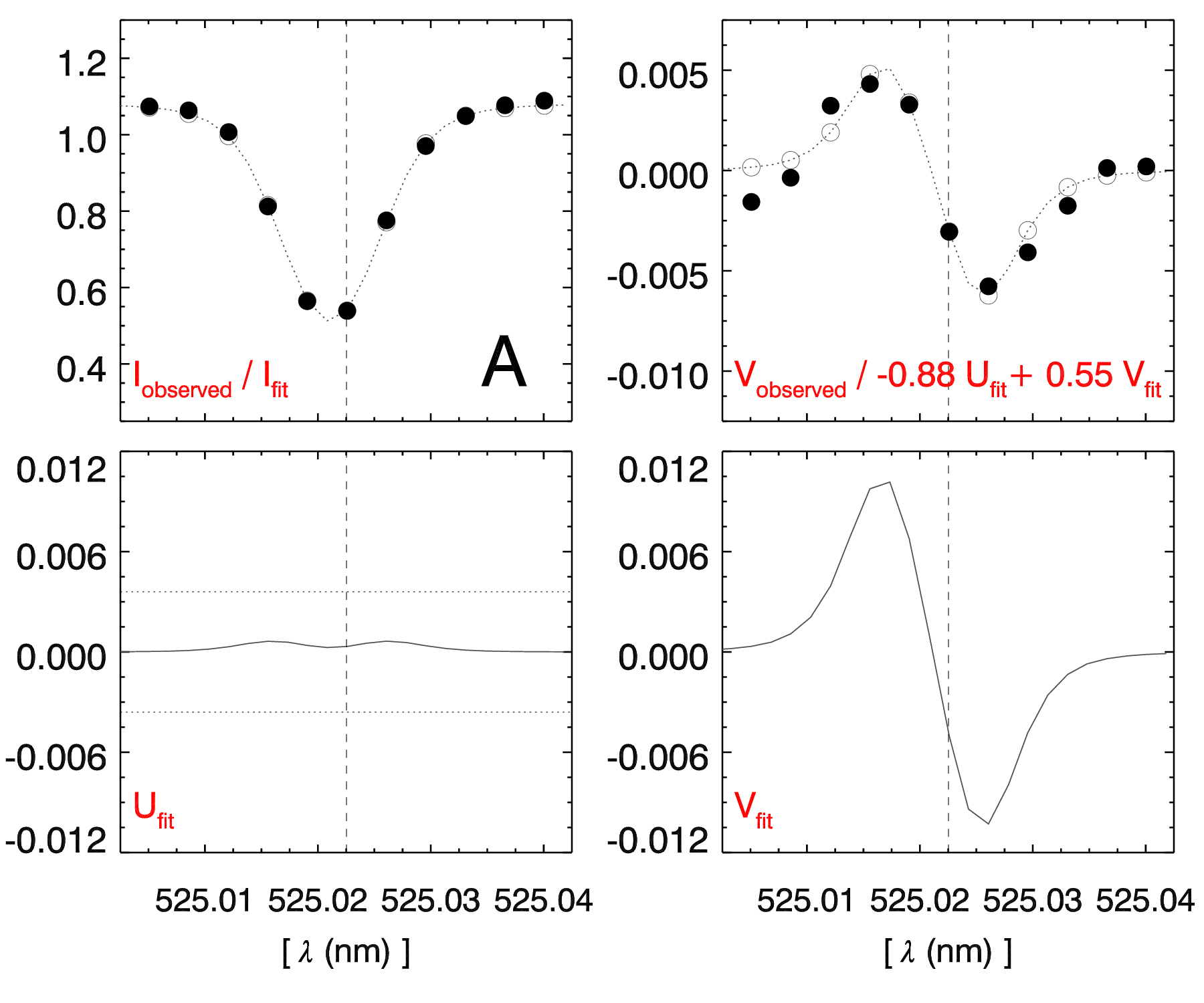}%
	\hspace{2pt}
	\includegraphics[trim=0 85 0 5, clip, scale=.1925]{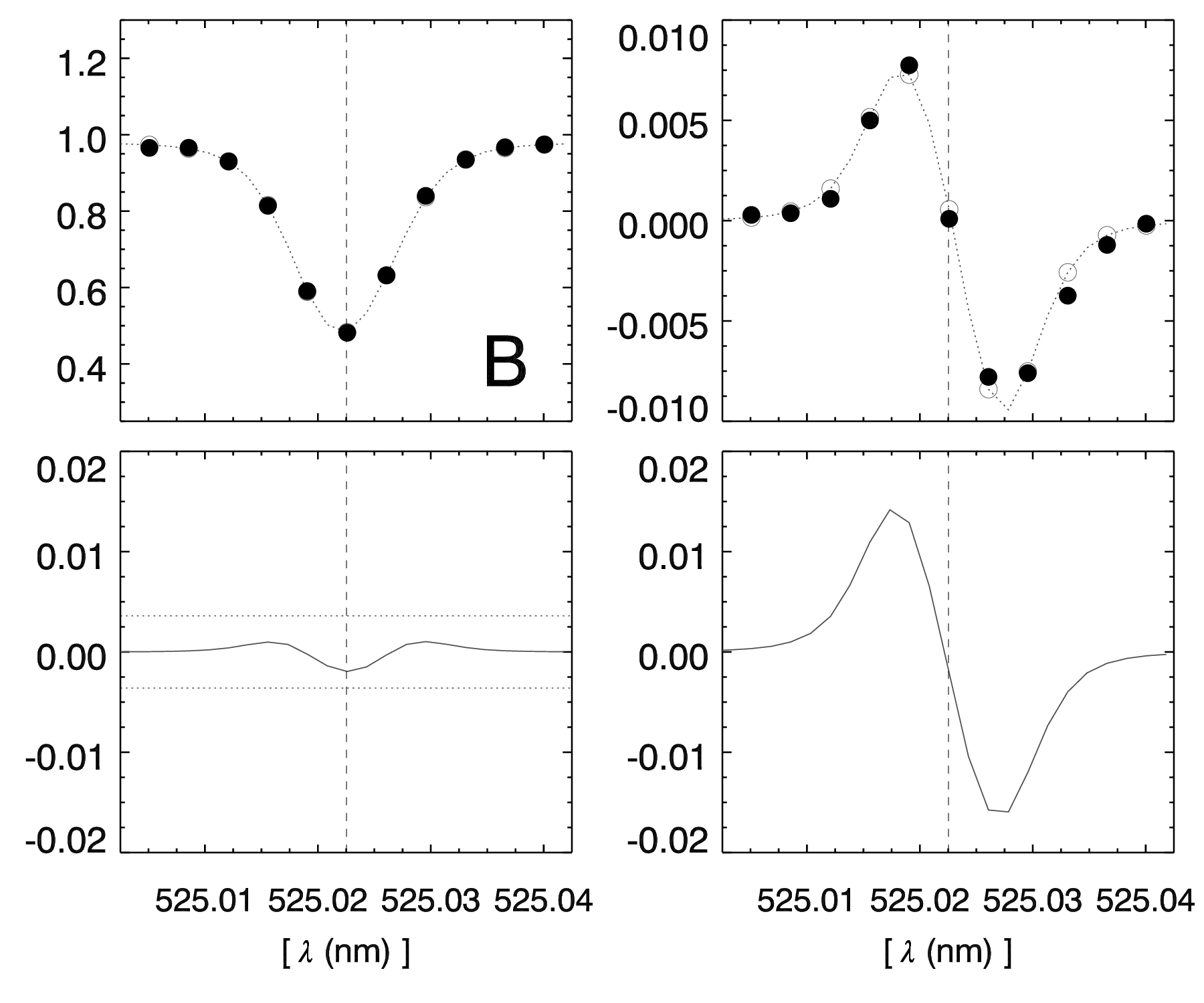}%
	\hspace{2pt}
	\includegraphics[trim=0 85 0 5, clip, scale=.1925]{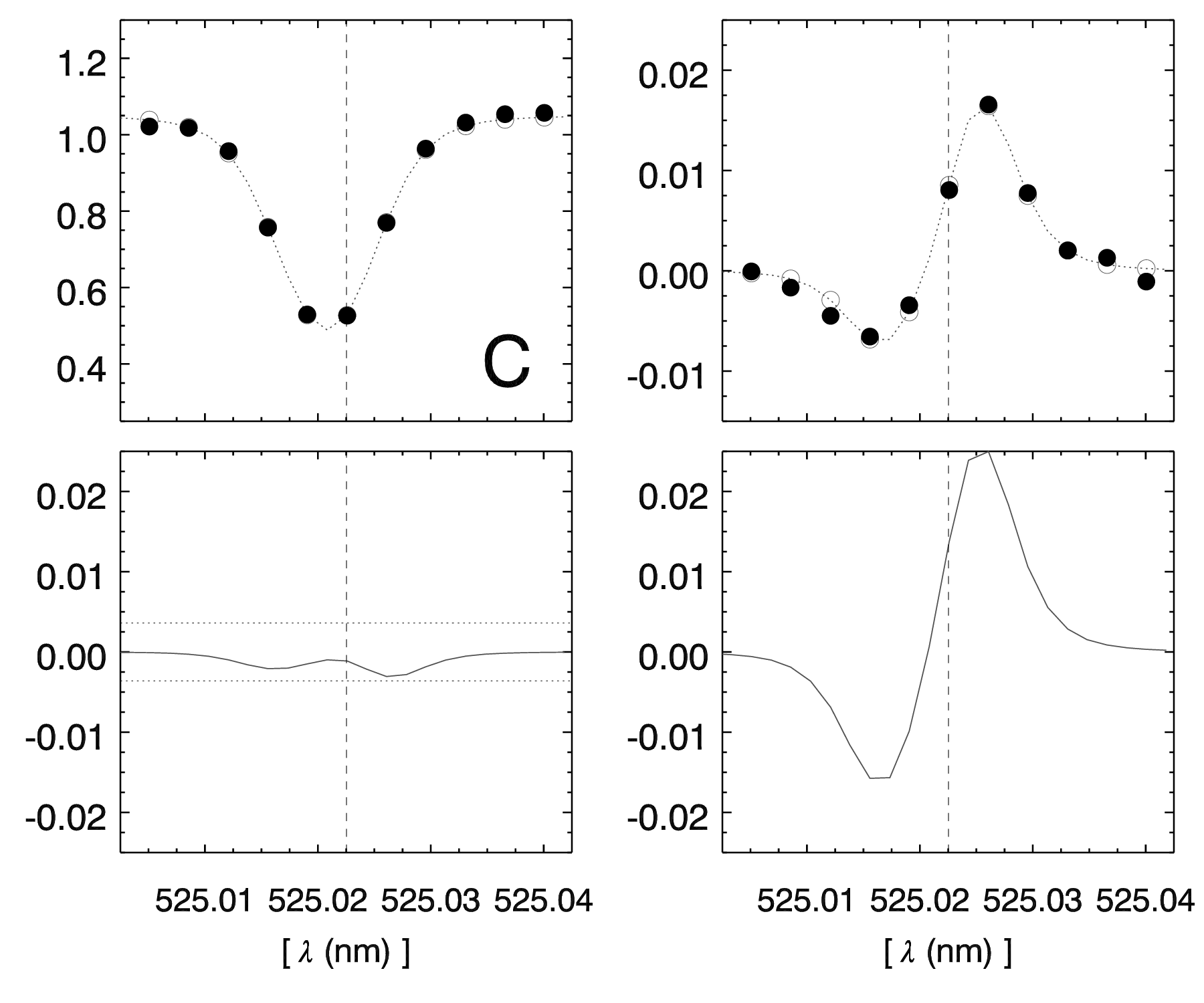}
	
	\vspace*{4pt}
	\includegraphics[trim=0 85 0 5, clip, scale=.1925]{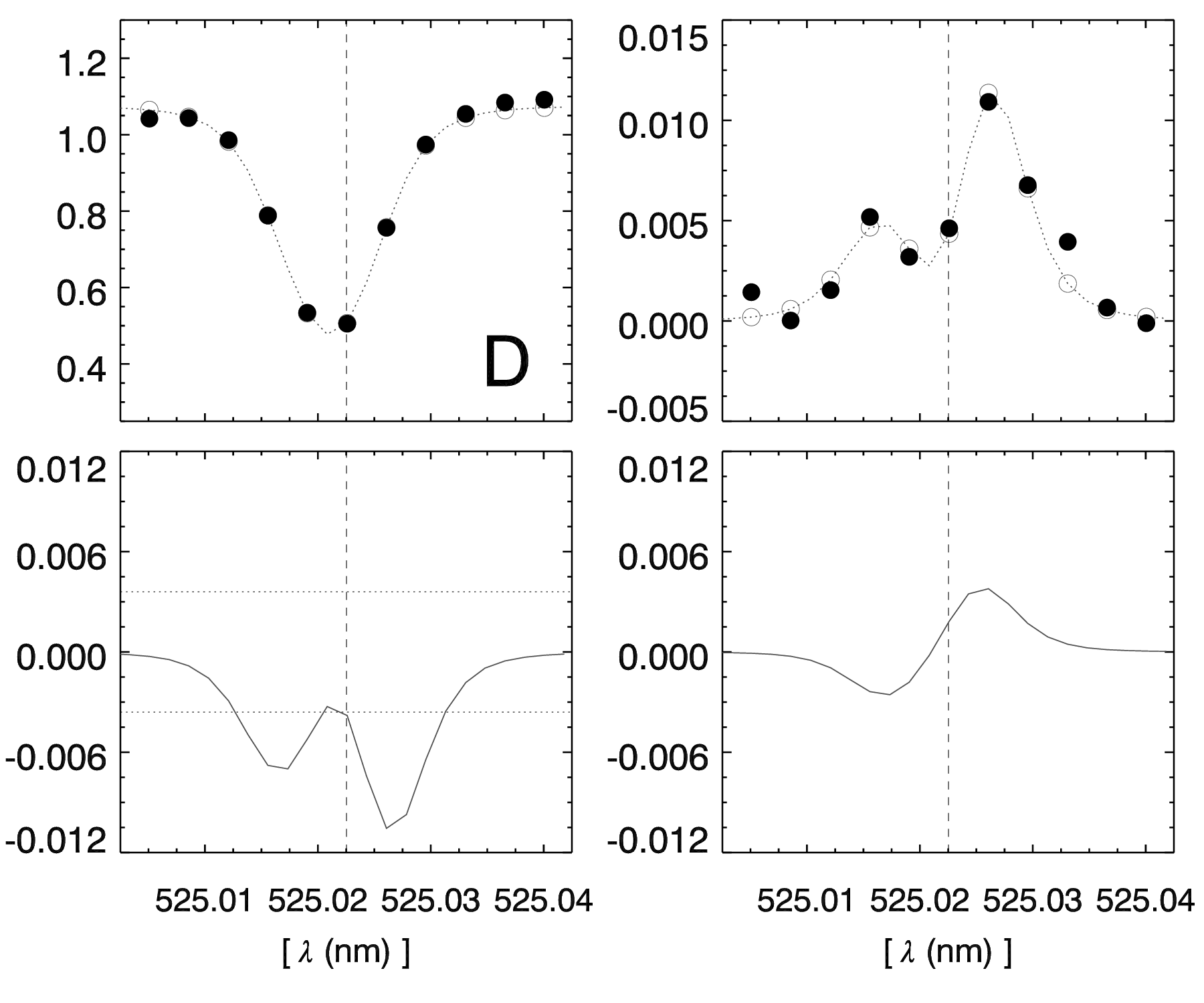}%
	\hspace{2pt}
	\includegraphics[trim=0 85 0 5, clip, scale=.1925]{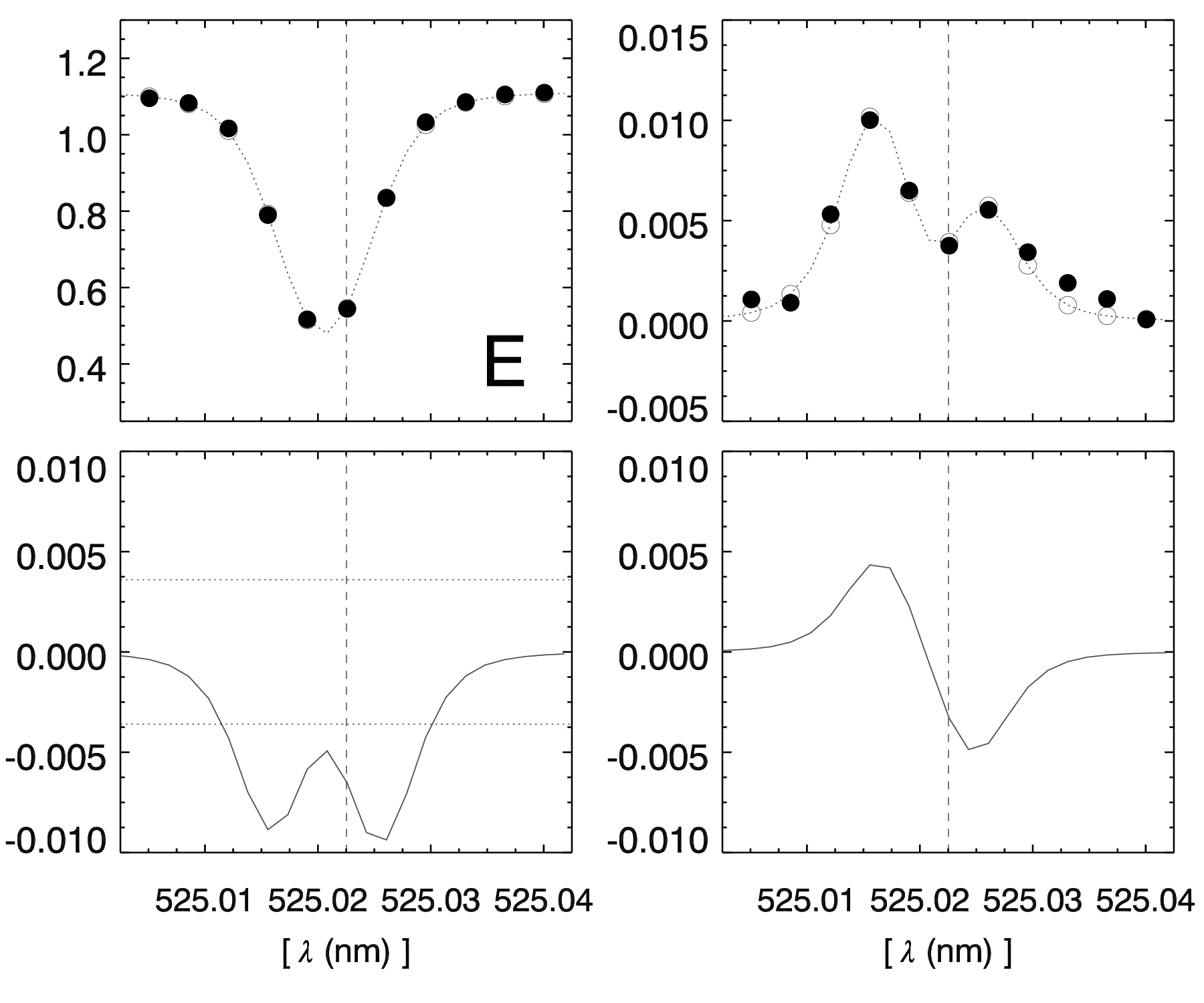}%
	\hspace{2pt}
	\includegraphics[trim=0 85 0 5, clip, scale=.1925]{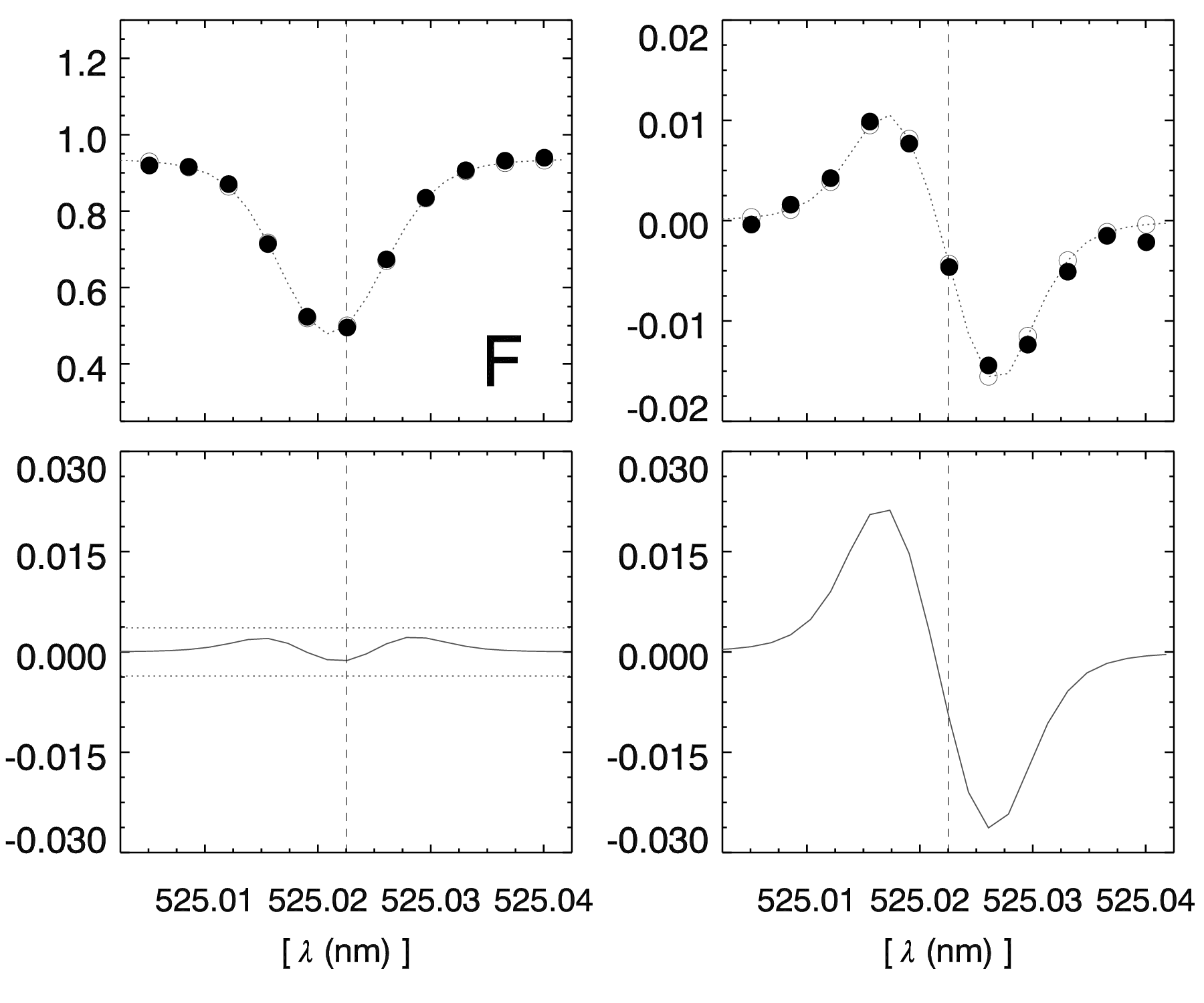}
	
	\vspace*{4pt}
	\includegraphics[trim=0 85 0 5, clip, scale=.1925]{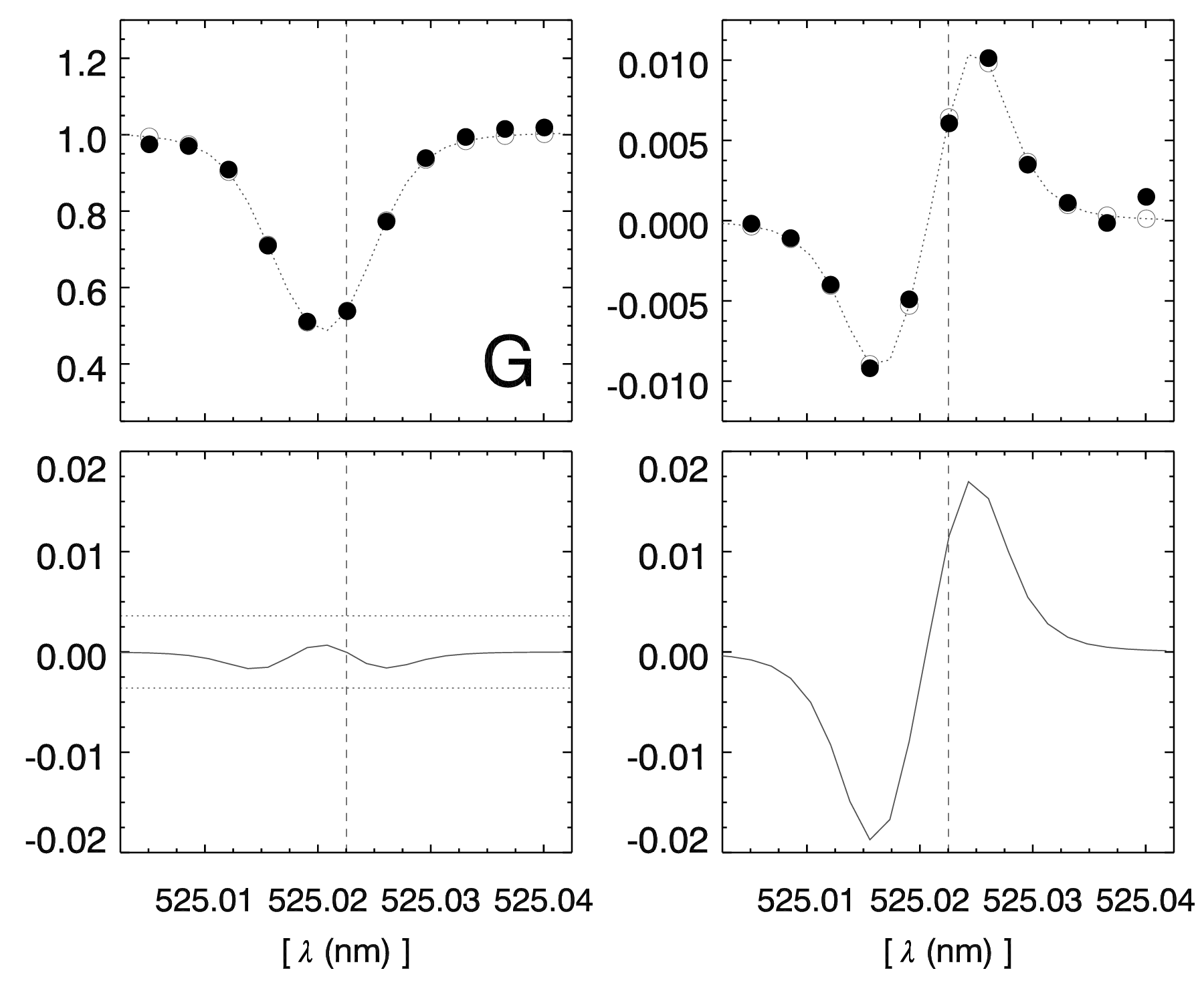}%
	\hspace{2pt}
	\includegraphics[trim=0 85 0 5, clip, scale=.1925]{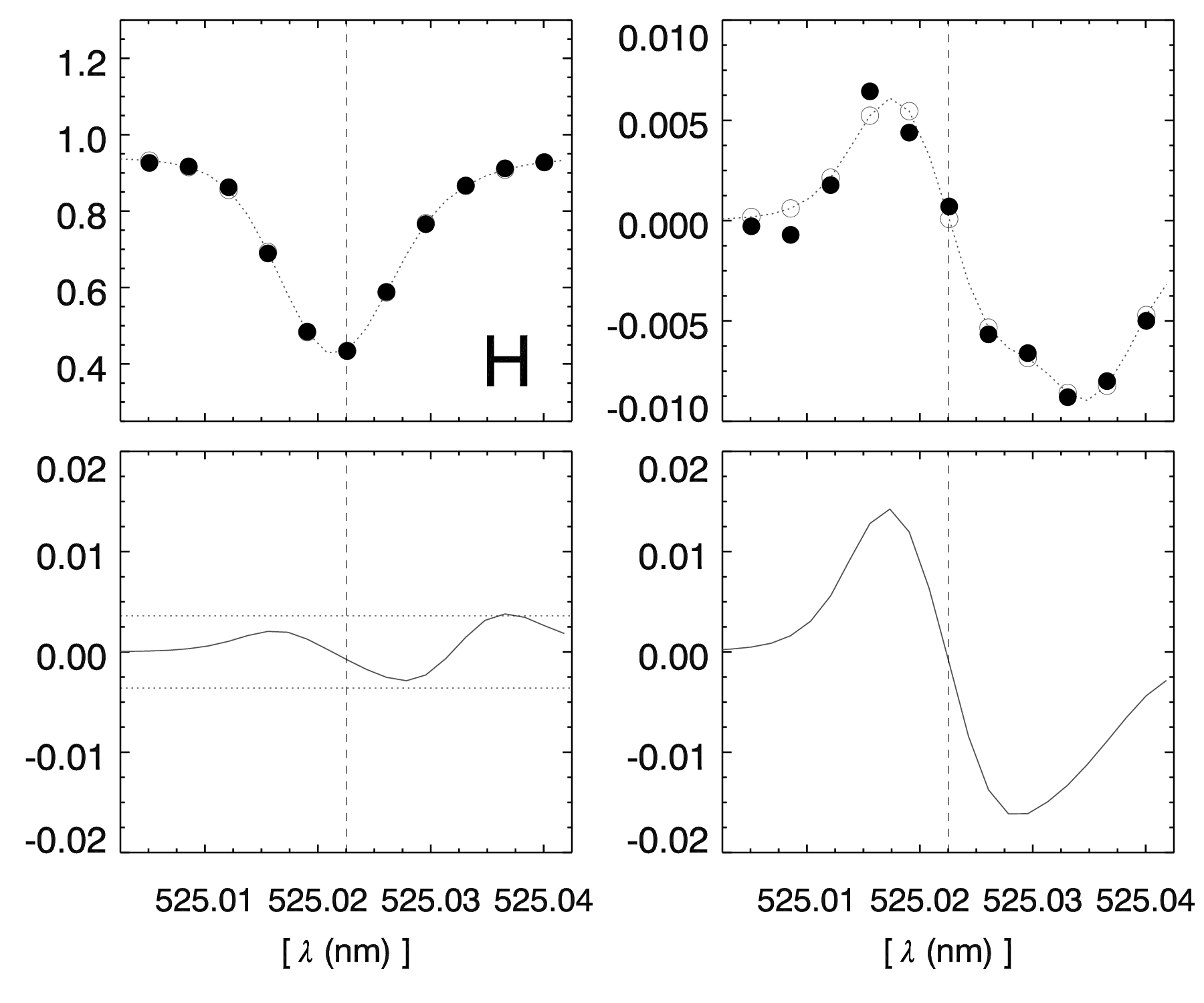}%
	\hspace{2pt}
	\includegraphics[trim=0 85 0 5, clip, scale=.1925]{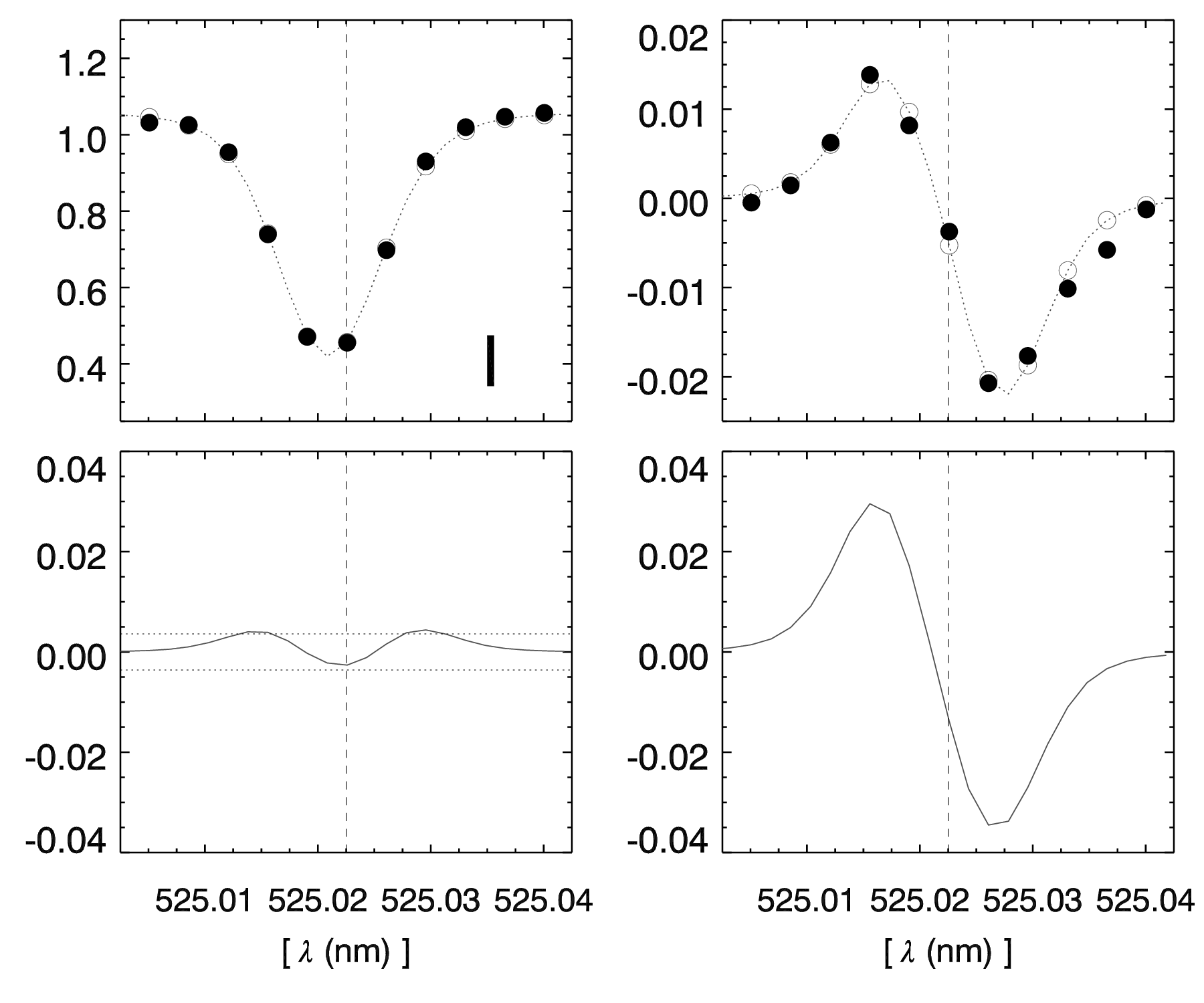}

	\vspace*{4pt}	
	\includegraphics[trim=0 15 0 5, clip, scale=.1925]{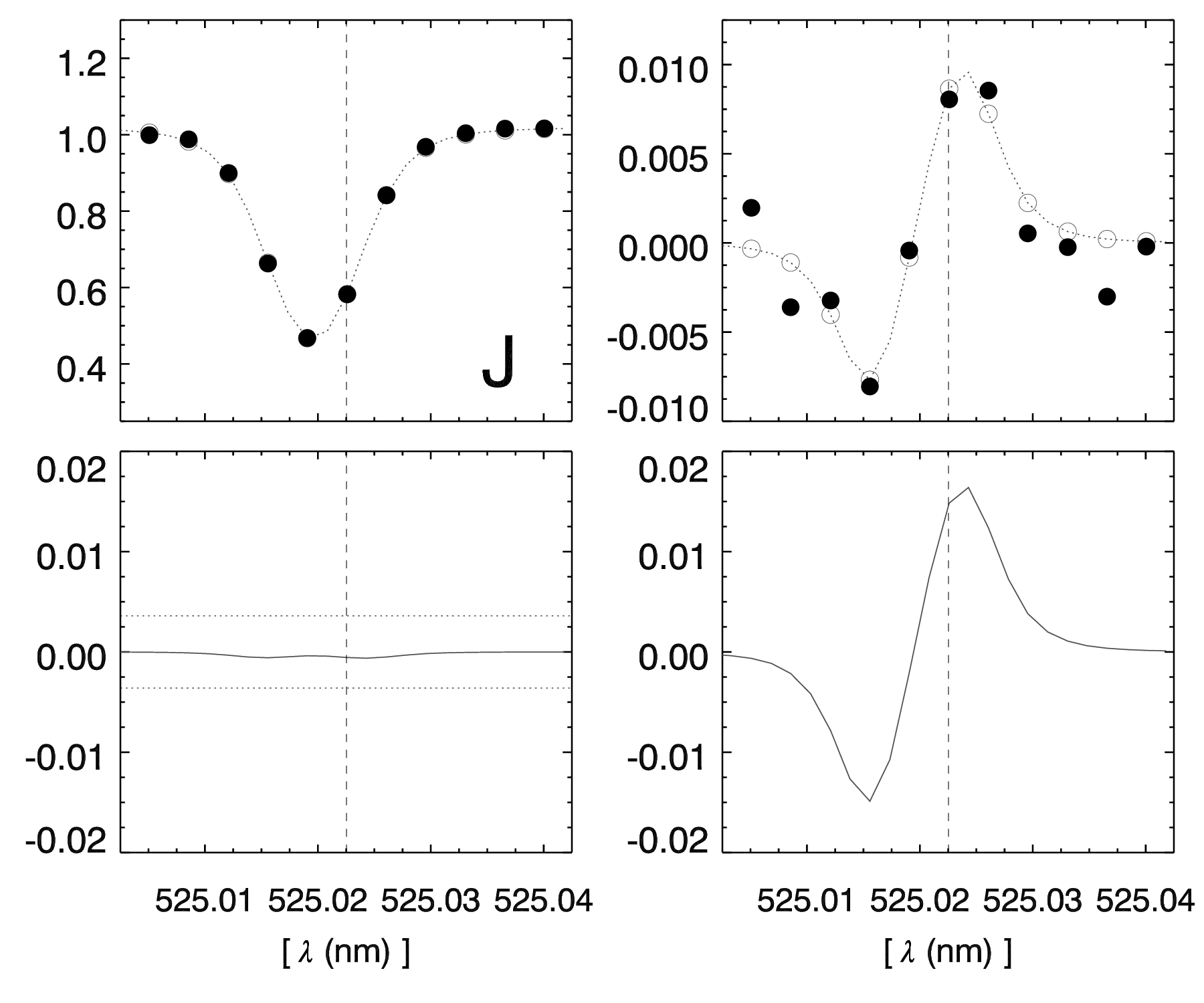}%
	\hspace{2pt}
	\includegraphics[trim=0 15 0 5, clip, scale=.1925]{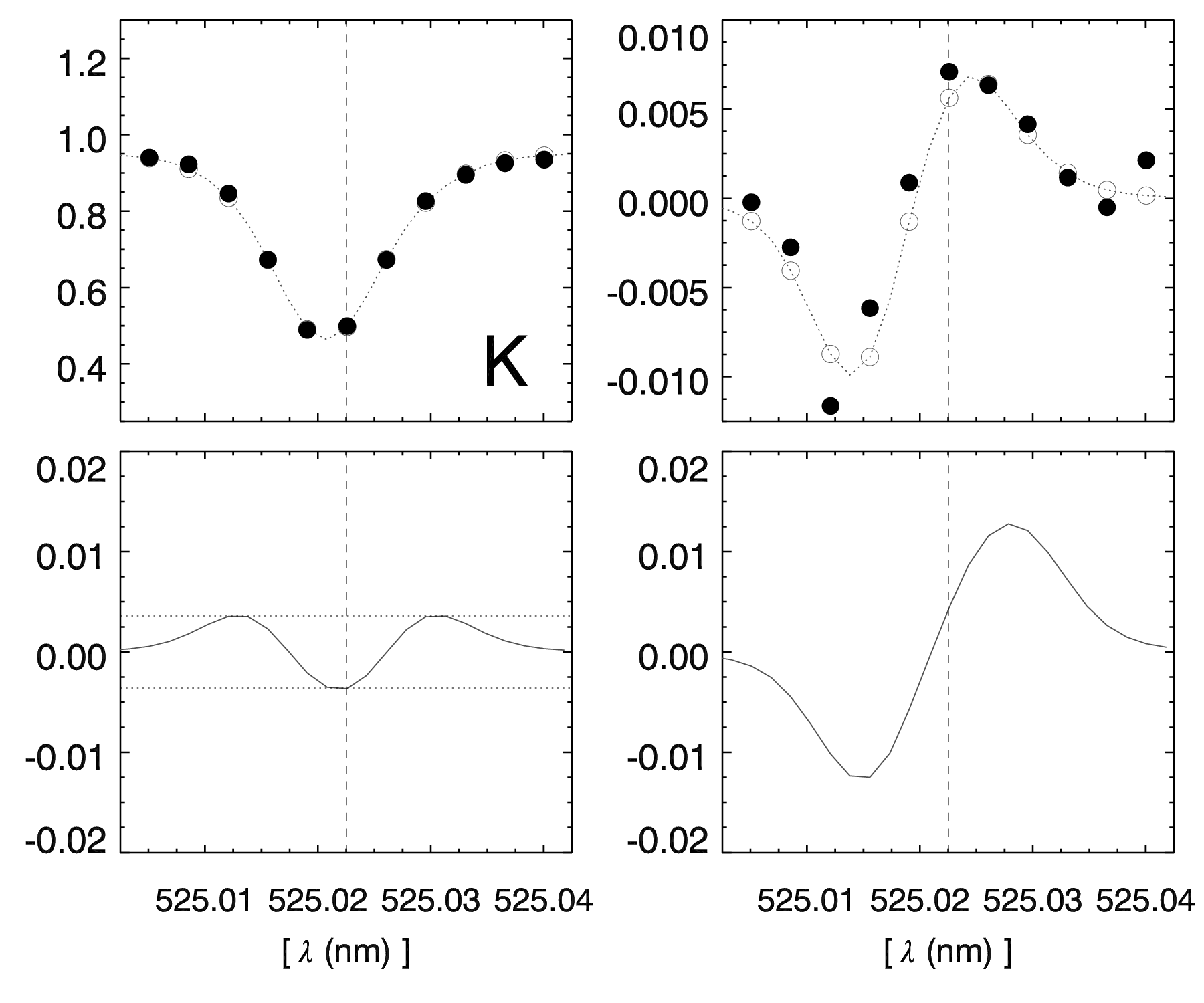}%
	\hspace{2pt}
	\includegraphics[trim=0 15 0 5, clip, scale=.1925]{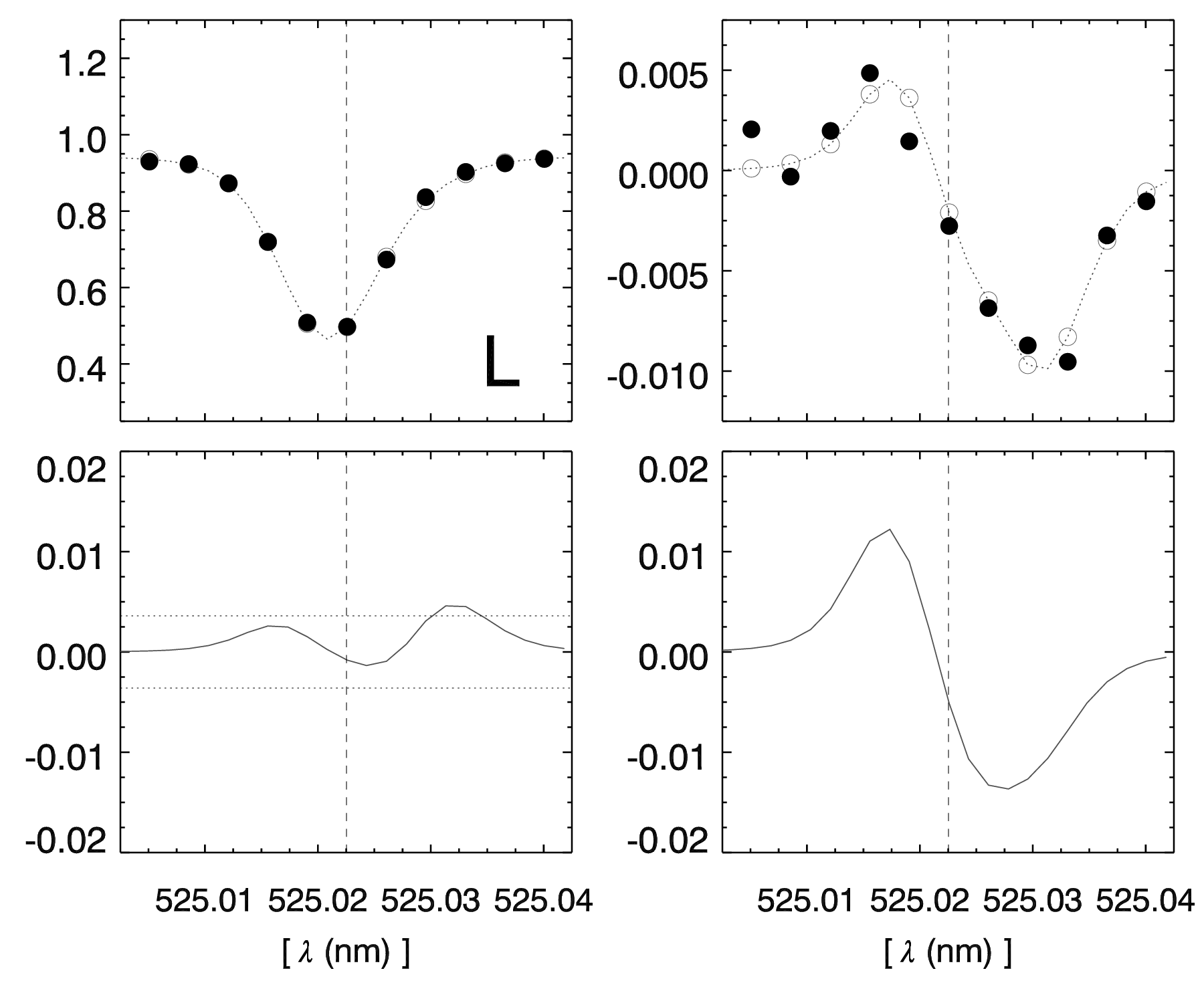}
	\caption{Stokes profiles for the positions indicated with capital letters in Figure~\ref{fig4}. A set of four panels represents the profiles at a given position. Thus, at each spatial position, we plot Stokes~\textit{I} (upper-left panel) and the linear combination represented by Eq.~\ref{eqUV} (upper-right panel): filled circles represent the observed data, empty circles represent the fitted values obtained from the \emph{SirUV} inversions. The dotted line is a spline interpolation of the fitted value. Furthermore, the graph displays the fitted profiles of Stokes~\textit{U} (lower-left panel) and Stokes~\textit{V} (lower-right panel). The dashed lines plotted over the Stokes~\textit{U} profiles indicate three times the value of the noise. \label{fig6}}
\end{figure*}

\subsection{Polarization signatures}

In this subsection we will analyze in detail the polarization signatures present across the EG, in order to characterize its magnetic configuration and to study their spatial distribution in the area of the EG. 

The crosses that are plotted over Figure~\ref{fig4} indicate individual pixels, whose Stokes profiles are shown in Figure~\ref{fig6}. In addition, the presence of these marks allows us to perceive some slight apparent displacements of their position between the $V_{\mathrm{integrated}}$ maps and the magnetic flux density maps. 
This is the case, for instance, for position H. It appears just in the middle between two patches of opposite polarity in the $V_{\mathrm{integrated}}$ map, while it appears to be located just inside the positive polarity in the magnetic flux density map. Similarly, the cross relevant to position L at the center of the four descendant granules also appears at the center of the intergranular lanes with positive polarity in the map of magnetic flux density, whereas it seems to be displaced to the bottom-left direction in the $V_{\mathrm{integrated}}$ map.
This is due to the fact that the Stokes~$V_{\mathrm{measured}}$ signal, and thus the $V_{\mathrm{integrated}}$ quantity, is altered by the small contribution brought by Stokes~$U_{\mathrm \sun}$.

\begin{figure}[!t]
	\centering
	\includegraphics[trim= 55 250 200 225, clip, scale=.57]{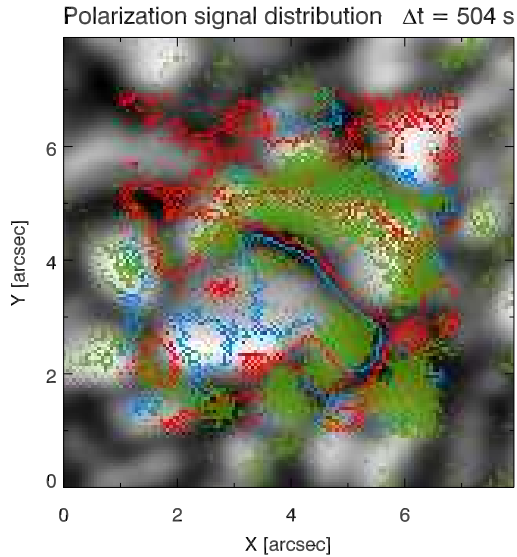}
	\caption{Distribution of the polarization signals across the EG at a given time ($\Delta t = 504$~s) during its evolution. Red (blue) contours enclose areas with $V_{\mathrm{fitted}}$ larger than three times the noise level, with positive (negative) polarity. Green regions indicate areas with $U_{\mathrm{fitted}}$ larger than than three times the noise level. \label{figpol}}
\end{figure}

In Figure~\ref{fig6} we plot the polarization profiles for each one of the twelve positions indicated in Figure~\ref{fig4}. In the upper panels of each set of four panels referring to a given solar location, we display the observed and fitted Stokes~\textit{I} and the Stokes~$V_{\mathrm{measured}}$ together with the linear combination described in Eq.~\ref{eqUV}, which is constructed with the superposition of the fitted Stokes~\textit{Q} and~\textit{V} parameters ($U_{\mathrm{fitted}}$ and~$V_{\mathrm{fitted}}$, respectively). Furthermore, in the lower panels we plot separately the Stokes~$U_{\mathrm{fitted}}$ and~$V_{\mathrm{fitted}}$ profiles emerging from the \emph{SirUV} inversions. 

In general, we can see that the agreement between the observed data and the fitted profiles is rather good, in particular for Stokes~\textit{I} profiles. Many $V_{\mathrm{measured}}$ profiles (positions A, B, F, G, J) are fitted with a strongly dominant Stokes~\textit{V} component, the Stokes~$U_{\mathrm{fitted}}$ signal lying entirely under three times the noise level. By contrast, some strongly asymmetric $V_{\mathrm{measured}}$ profiles are well understood in terms of superposition of Stokes~$V_{\mathrm{fitted}}$ with a significant Stokes~$U_{\mathrm{fitted}}$ signal. This is clearly the case of the profiles relevant to positions D and E, and to a lesser extent of those found at positions H and L. The profiles observed at positions C, I, and K represent an intermediate case, as the Stokes~$U_{\mathrm{fitted}}$ signal required to fit the measured data just reaches above the noise threshold.      

These examples also show that, indeed, the \emph{SirUV} code mainly assigns the wavelength symmetric part of the Stokes~$V_{\mathrm{measured}}$ profile to $U_{\mathrm{fitted}}$ and the antisymmetric part to $V_{\mathrm{fitted}}$. This is the case even though gradients are allowed in $v_{\mathrm LOS}$ and $B$ (they have a linear dependence with $\tau$ in the inversion), so that in principle asymmetric Stokes~$V$ profiles can be generated \citep[e.g.,][]{Illing:75,Schussler:88,Solanki:89}. Nonetheless, they might be interpreted in terms of a transversal component of the field, i.e., requiring a significant Stokes~$U_{\mathrm{fitted}}$ signal to fit the Stokes~$V_{\mathrm{measured}}$ profiles. This happens as the code is supplied with only two independent Stokes parameters and, therefore, being not able to determine the inclination, cannot discriminate between the asymmetry introduced by LOS gradients or the contribution of Stokes~$U$ signals generated by a transversal component.

\begin{figure*}[t]
	\centering
	\includegraphics[trim= 10 470 280 10, clip, scale=.57]{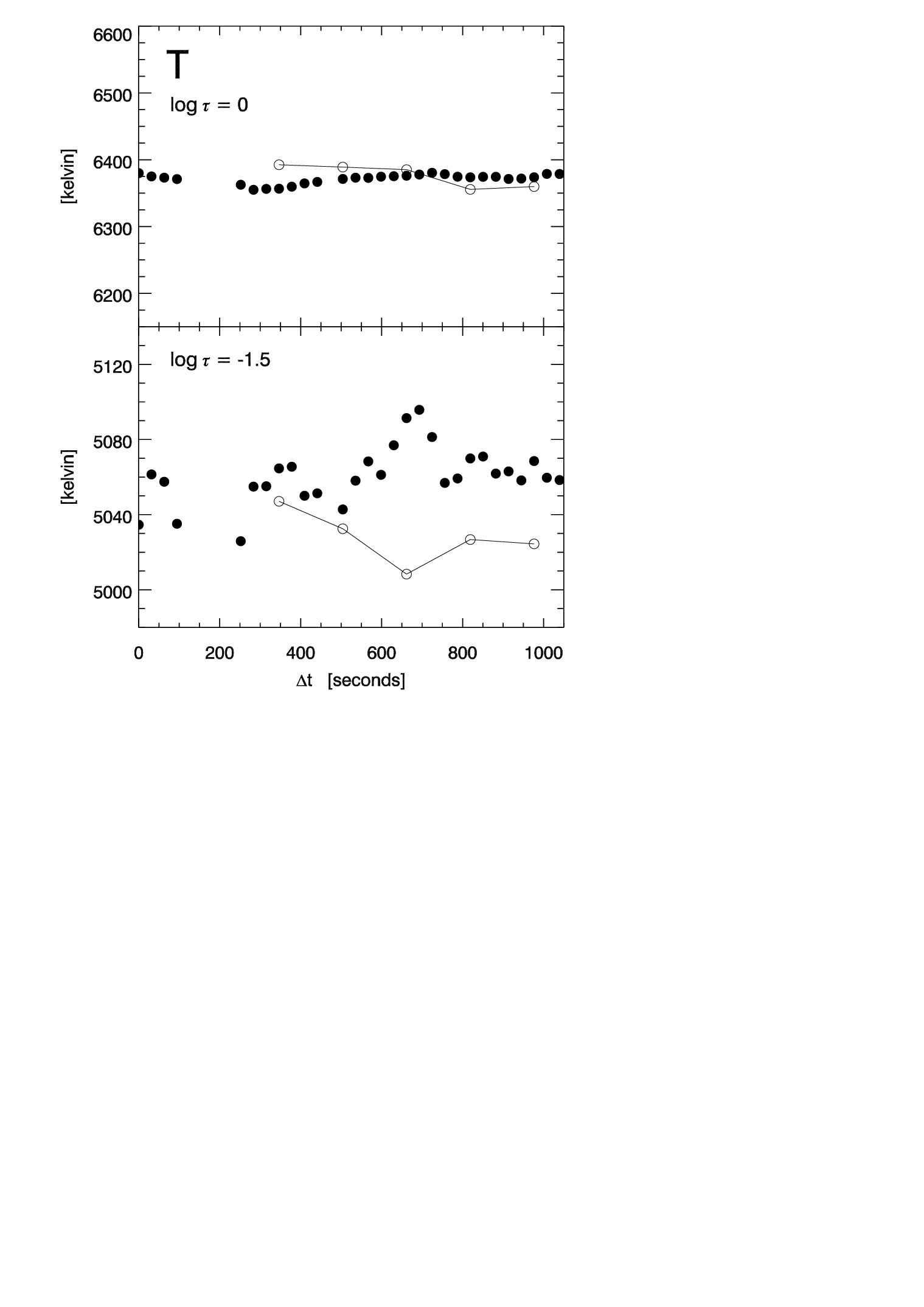}%
	\includegraphics[trim=105 470 280 10, clip, scale=.57]{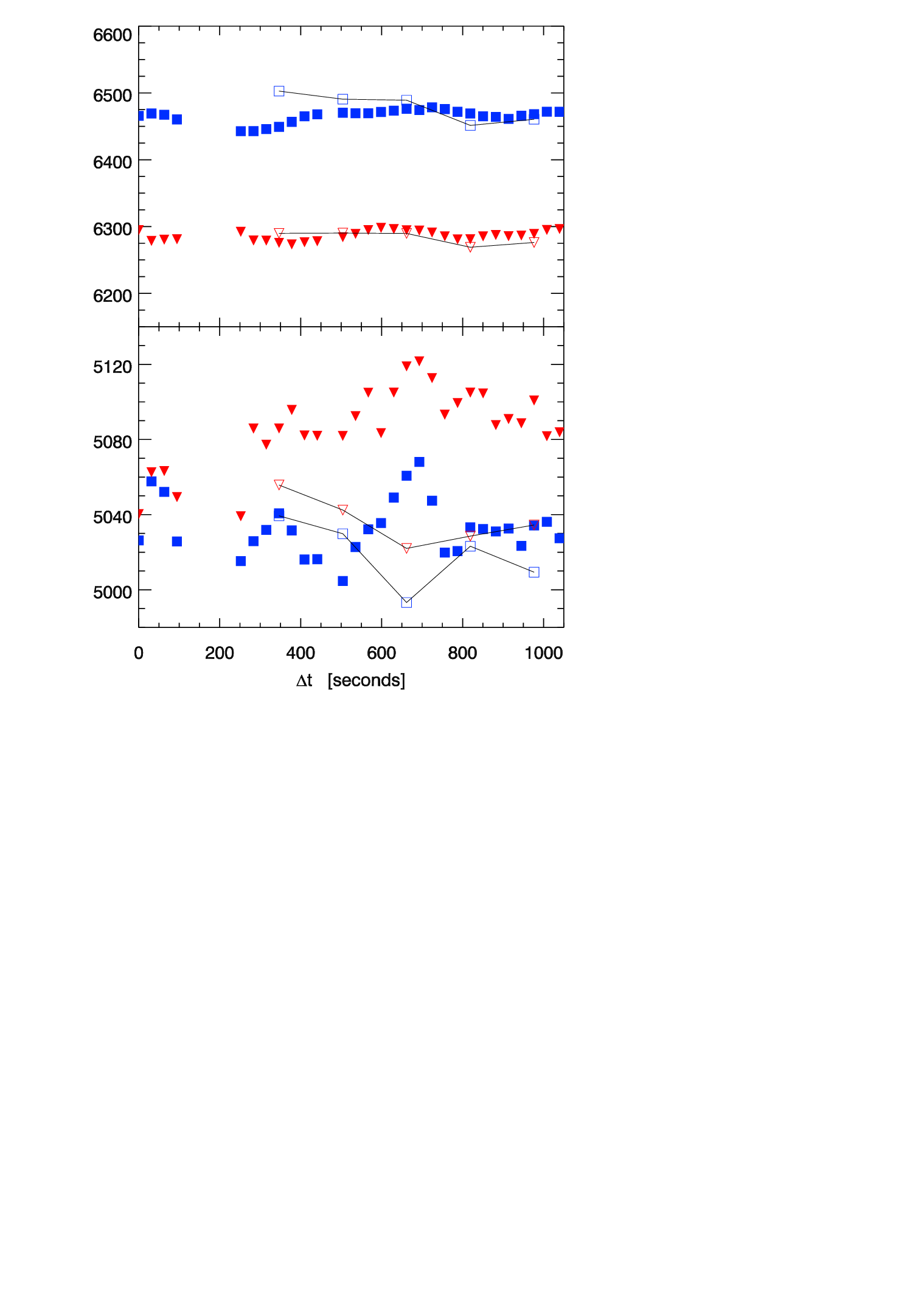}	
	\caption{\textit{Left panels}: Evolution of the temperature (black circles) averaged within FoV$_{EG}$ (solid box in Figure~\ref{fig0}), and of $T$ (empty circles) averaged within FoV$_{QS}$ (dashed box in Figure~\ref{fig0}), at $\log \tau =0$ (top) and $\log \tau =-1.5$ (bottom). \textit{Right panels}: Same, for granular (squares, blue) and intergranular (triangles, red) regions, respectively, present in both subFoVs. \label{fig7}}
\end{figure*}

As can be seen in Figure~\ref{figpol}, the distribution of the Stokes signals across the EG clearly indicates the presence of horizontal fields in the feature. In fact, some regions with Stokes~$U_{\mathrm{fitted}}$ signals above three times the noise level, shown in green color, are located in bright parts of the EG. Others are found next to neutral lines, close to the edges of adjacent red/blue contours, which indicate regions with Stokes~$V_{\mathrm{fitted}}$ signal larger than three times the noise level with positive (negative) polarity. In particular, large $U_{\mathrm{fitted}}$ signals are found around the neutral line, being located at about $5\arcsec \times \left[4\arcsec, 6\arcsec\right]$.

\subsection{Thermodynamical parameters}

The capabilities of the \emph{SirUV} code, which includes full radiative transfer, allow us to study the thermodynamical parameters in the region of the solar atmosphere where flux emergence is taking place. In order to compare the thermal and pressure stratification in the EG containing the emerging flux region with those found in the undisturbed photosphere, we have studied the thermodynamical properties in FoV$_{EG}$ and FoV$_{QS}$ (see Figure~\ref{fig0}). 

In Figure~\ref{fig7} we plot the evolution of the average values of $T$ at two optical depths ($\log \tau =0$ and $\log \tau =-1.5$) for the entire subFoVs (left panels) and for granular (blue) and intergranular regions (red) separately (right panels). Note that before $\Delta t \approx 250$ the very quiet Sun region in FoV$_{QS}$ was outside the IMaX full FoV.

\begin{figure*}[!t]
	\centering
	\includegraphics[trim= 10 235 280 10, clip, scale=.57]{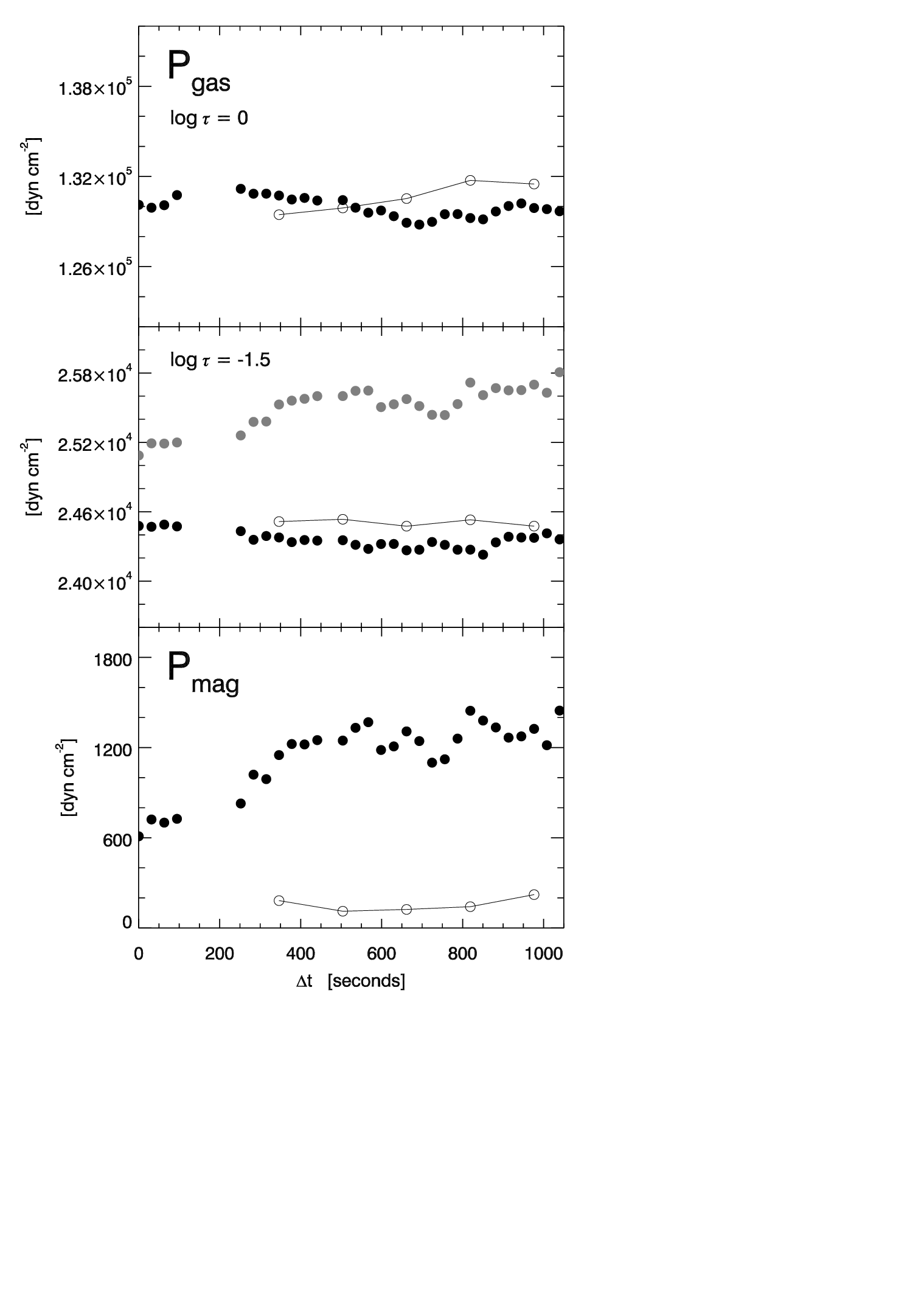}%
	\includegraphics[trim=105 235 280 10, clip, scale=.57]{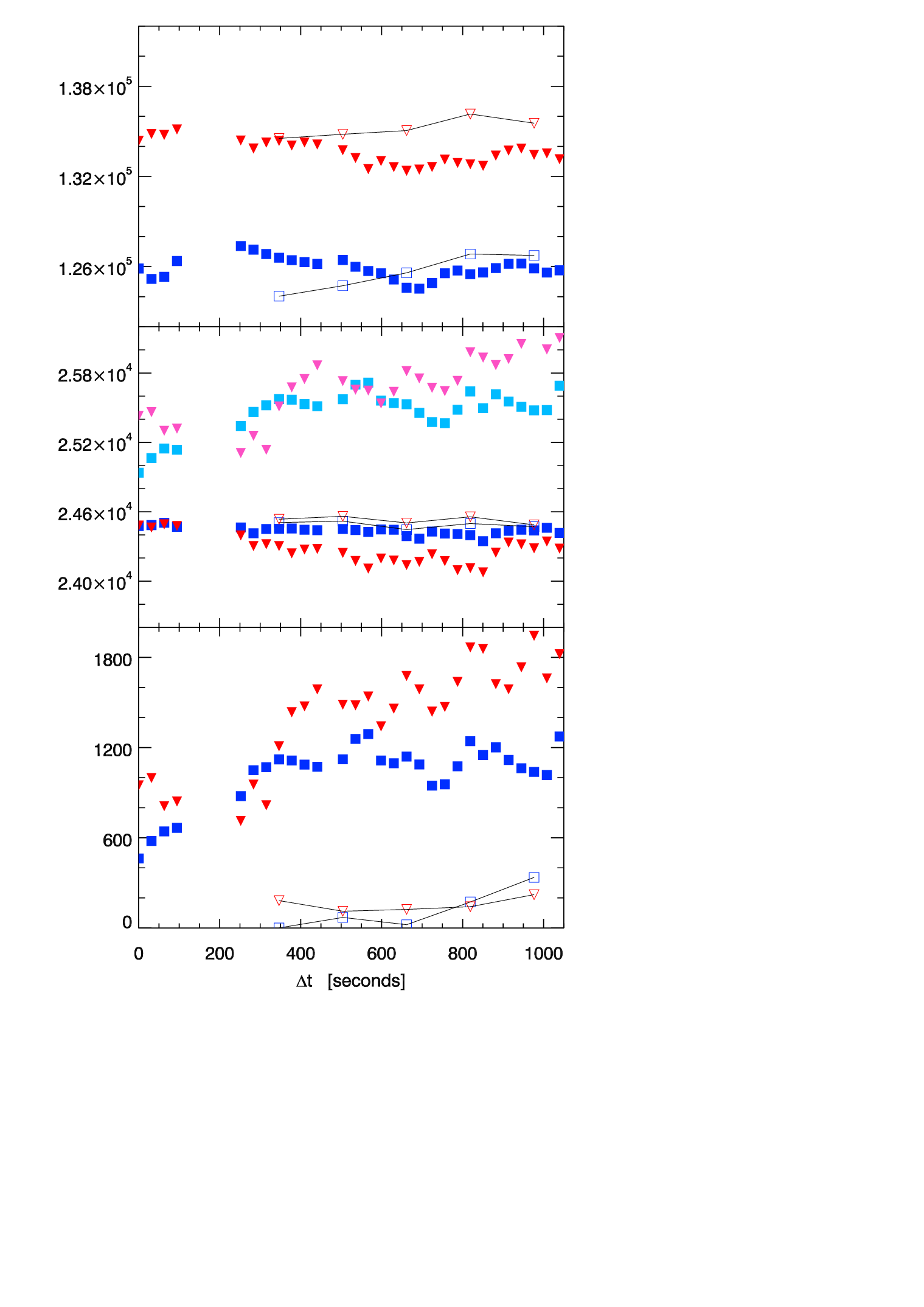}	
	\caption{\textit{Left panels}: Evolution of the gas pressure (black circles) and of total pressure (grey circles) averaged within FoV$_{EG}$ (solid box in Figure~\ref{fig0}), and of $P_{\mathrm {gas}}$ (empty circles) averaged within FoV$_{QS}$ (dashed box in Figure~\ref{fig0}), at $\log \tau =0$ (top) and $\log \tau =-1.5$ (middle). The trend of $P_{\mathrm {mag}}$ is also shown (bottom), for FoV$_{EG}$ (black circles) and FoV$_{QS}$ (empty circles). \textit{Right panels}: Same, for granular (squares, blue color scheme) and intergranular (triangles, red color scheme) regions, respectively, present in both subFoVs. In the middle right panel light blue squares and pink triangles refer to the total pressure. \label{fig8}}
\end{figure*}

The graphs indicate that the temperature at $\log \tau =0$ is basically the same in the emerging flux region and in the very quiet Sun, except for a modestly lower value in the intergranular lanes in FoV$_{QS}$. Instead, at $\log \tau =-1.5$, we find an increase of $T$ in FoV$_{EG}$ during the evolution of the emerging flux region, with a subsequent, smaller decrease after $\approx 700$~s. There is also a difference with respect to the very quiet region: the maximum spread is around $100$~K in intergranular regions, whereas it has a lower value in granules. 

\begin{figure}[!t]
	\centering
	\includegraphics[trim= 10 720 280 10, clip, scale=.57]{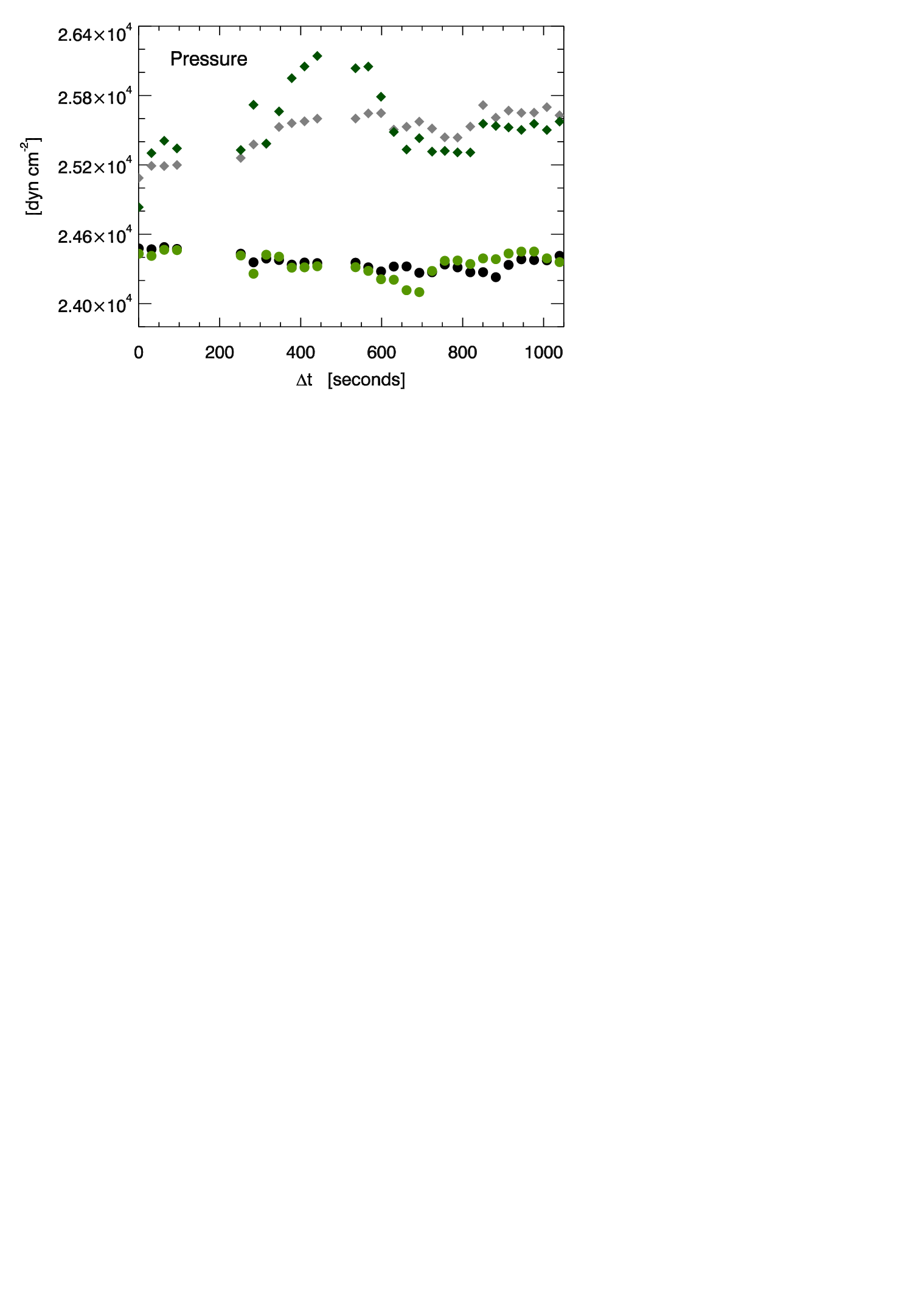}	
	\caption{Evolution of the gas pressure (circles) and of total pressure (diamonds) averaged within the entire FoV$_{EG}$ (black/grey colors) and within the region of interest occupied by the EG inside FoV$_{EG}$ (light green/dark green colors). \label{fig8extra}}
\end{figure}

In Figure~\ref{fig8} (left panels) we show the trend of the gas pressure averaged in FoV$_{EG}$ and FoV$_{QS}$ at two optical depths ($\log \tau =0$ and $\log \tau =-1.5$). The magnetic and total (gaseous + magnetic at $\log \tau =-1.5$) pressure values in the EG region are also displayed. 

The plots in Figure~\ref{fig8} (left panels) exhibit some fluctuations of $P_{\mathrm {gas}}$ with time. The rms variations of $P_{\mathrm {gas}}$ are about 5\% for $\log \tau =0$ and about 3\% for $\log \tau =-1.5$.  Furthermore, they reveal that magnetic pressure, due to the emerging field, 
increases the total pressure by about 5\% at $\log \tau =-1.5$. 
This seems to be in agreement with the larger pressure expected in EGs for their expansion. 
By contrast, the pressure in the very quiet Sun region has a stationary trend.

Figure~\ref{fig8} (right panels) shows the same quantities, for granules (blue color scheme) and intergranular lanes (red color scheme) separately. Both granular and intergranular regions are affected by the presence of the magnetic field, even if the effect on intergranules is slightly more pronounced. In these areas we also observe larger fluctuations with time. Again in these graphs, at $\log \tau =-1.5$ we find an increase of the total pressure, which is stronger in intergranular lanes. 

In Figure~\ref{fig8extra} we compare the gaseous and the total pressure for the entire FoV$_{EG}$ and for the region of interest occupied by the EG within the same subFoV, defined by the hand-drawn contours that have been already used to derive the plots shown in Figure~\ref{fig5}. The graph in Figure~\ref{fig8extra} illustrates that, during the first 400~s, the total pressure is more enhanced mainly above and directly around the EG than in the neighbouring granules contained in FoV$_{EG}$.

Note that $T$, $P_{\mathrm {gas}}$, and $P_{\mathrm {mag}}$ are shown at a given optical depth $\tau$ for each pixel. However, that $\tau$ refers to very different values of geometrical height $z$ at different times or in different pixels that have different Wilson depressions, due to the presence of the magnetic field. In particular, this occurs in areas with distinct properties such as granules and intergranular lanes. 

Therefore, to study such effects, we have compared our results with a numerical model, checking how strongly the $\log \tau =0$ and $\log \tau =-1.5$ surfaces move up and down in such a datacube and how much the pressure varies just due to these height changes. We have used a snapshot from a three-dimensional non-ideal compressible radiation MHD simulation calculated with the MURaM code \citep{Vogler:05}, with a unipolar seed of  
homogeneous vertical magnetic field of $30$~G representing the quiet Sun. The data cubes cover $6$~Mm in both horizontal directions, with a cell size of about $10.4$~km (0\farcs14 $\mathrm{px}^{-1}$), while in the vertical direction they extend $1.4$~Mm with a $14$~km cell size \citep[for further details, see][]{Tino:14}.

On average, $\log \tau =0$ and $\log \tau =-1.5$ for the continuum at $500$~nm are reached about $z_1=875$~km and $z_2=1100$~km above the bottom boundary, respectively. The rms variations of $z$ are $\sigma_{z_1}=38$~km for $\log \tau =0$ and $\sigma_{z_2}=34$~km for $\log \tau =-1.5$. 

At fixed $\log \tau$, the rms variations of $P_{\mathrm {gas}}$ are about 10\% for $\log \tau =0$ and about 9\% for $\log \tau =-1.5$. Conversely, at geometrical heights $z_1$ and $z_2$, the rms variations of $P_{\mathrm {gas}}$ are about 18\% and 24\%, respectively. If $P_{\mathrm {gas}}$ is evaluated at $z_1 + \sigma_{z_1}$ and $z_1 - \sigma_{z_1}$ it changes by a slightly larger factor of 20\%; similarly, $P_{\mathrm {gas}}$ changes by a factor of 30\% when it is evaluated at $z_2 + \sigma_{z_2}$ and $z_2 - \sigma_{z_2}$. Accordingly, the fluctuations and the slight decrease of pressure with respect to the very quiet Sun that we observe during the evolution of the EG might be of stochastic origin.

Nevertheless, to further explore this issue, we have studied the horizontal components of the gradient of pressure. First, we have computed these values at fixed optical depths, $\log \tau =0$ and $\log \tau =-1.5$. Then, we have calculated the gradients at fixed geometrical heights, those corresponding to the former values of $\tau$, i.e., $875$~km and $1100$~km, respectively. At $\log \tau =0$ the change in the module of gradients between isosurfaces of $\log \tau$ and of $z_1$ is relatively modest, leading to slightly smaller gradients at iso-$\tau$. Interestingly, this implies that the pressure variation being induced by the geometrical shift of the $\log \tau =0$ layer causes a decrease in the pressure gradient, likely due to the strong dependence of the optical depth on the gas density. That is supported by the fact that the rms of $P_{\mathrm {gas}}$ is smaller at iso-$\tau$ than at iso-$z$. At $\log \tau =-1.5$, the difference between iso-$z_2$ and iso-$\tau$ is larger, with significant smaller gradients at iso-$\tau$.

\begin{figure}[!b]
	\centering
	\includegraphics[trim= 20 20 20 90, clip, scale=.45]{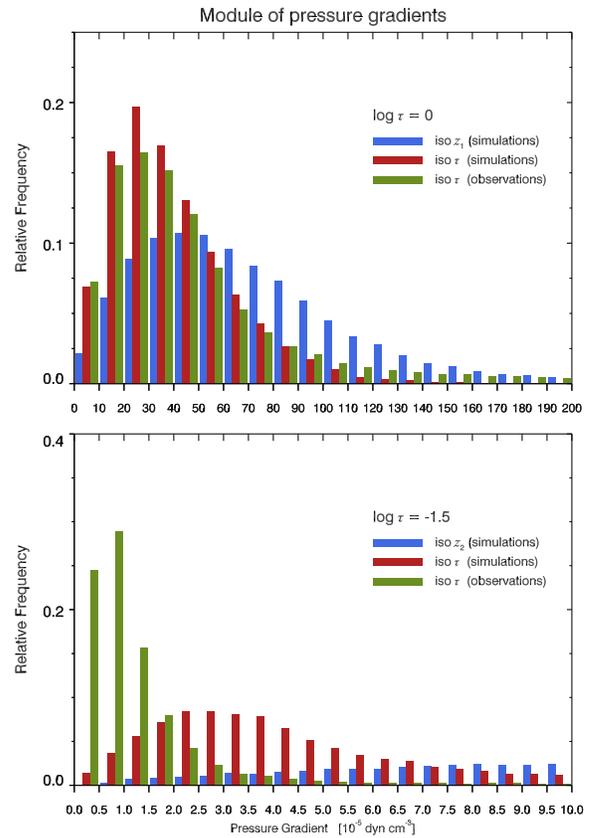}	
	\caption{\textit{Top panel}: Histograms of the modulus of the horizontal gradients of the pressure, for $\log \tau =0$, obtained at iso-$\tau$ and iso-$z$ in degraded simulations and at iso-$\tau$ in observations in FoV$_{EG}$. \textit{Bottom panel}: Same, for $\log \tau =-1.5$. \label{figmhd}}
\end{figure}

\begin{figure}[!t]
	\centering
	\includegraphics[trim= 20 490 20 0, clip, scale=.45]{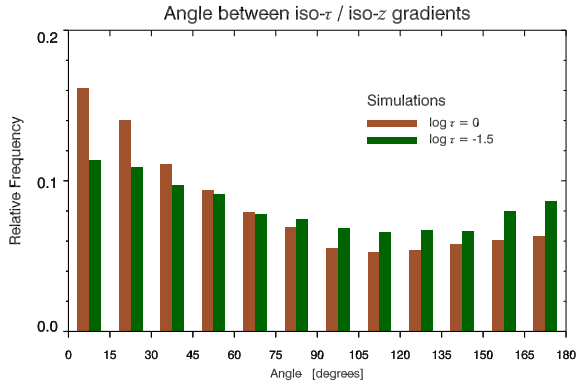}	
	\caption{Angle between horizontal gradients of the pressure at fixed optical depth and at fixed geometrical height in the degraded simulations, relevant to $\log \tau =0$ and $\log \tau =-1.5$. \label{figmhd2}}
\end{figure}

We have also evaluated the angle between the horizontal gradients at iso-$z$ and iso-$\tau$, to find if the signs of such gradients are preserved. In general, the majority of signs is preserved (i.e., angle $< 90^{\circ}$), in particular at $\log \tau =0$. Thus, at this height about 25\% of the pixels have gradients that differ by angles between $0^{\circ}$ and $15^{\circ}$, while for about 72\% of the pixels the gradients at equal $\tau$ and at equal $z$ have the same sign. At $\log \tau =-1.5$, the gradients differ by angles between $0^{\circ}$ and $15^{\circ}$ in about 17\% of the pixels, with the gradients having the same sign in about 65\% of all the pixels.

In order to compare these results with our observations, we have degraded the simulations, making a pixel 4 times larger, to be comparable to the spatial size of the pixel of the IMaX measurements. We have further convolved the pressure with a Gaussian with a FWHM of about 0\farcs3, to mimic the spatial resolution of the observations, and finally we have derived the gradients. 

In general, we find smoother and smaller variations in the degraded simulations. At iso-$\tau$, the rms variations of the degraded $P_{\mathrm {gas}}$ are about 6.5\% for $\log \tau =0$ and about 5\% for $\log \tau =-1.5$. At iso-$z_1$ and iso-$z_2$, the rms variations of $P_{\mathrm {gas}}$ are about 7\% and 12\%, respectively. Conversely, if $P_{\mathrm {gas}}$ is evaluated at $z_1 \pm \sigma_{z_1}$ and $z_2 \pm \sigma_{z_2}$, it changes by 20\% and 30\%, similar to the original simulations.

In Figure~\ref{figmhd} we compare the histograms of the relative frequency of the values found in the degraded simulations, at iso-$\tau$ and iso-$z$, and in the observations, at fixed $\log \tau$. At $\log \tau =0$, the change between iso-$z_1$ and iso-$\tau$ in simulations is still modest, even if slightly larger than in the case of original simulations. By contrast, there is substantial agreement between the modulus of the horizontal gradients at iso-$\tau$ in simulations and observations. Hence, when the different spatial resolution is properly taken into account, the histograms relevant to the degraded simulations look more similar to the observations. At $\log \tau =-1.5$, the difference between iso-$z_2$ and iso-$\tau$ in simulations is by far larger and ever smaller gradients are found in IMaX observations. This difference may have to do with problems in the simulations, although it could be also caused by the fact that SIR assumes hydrostatic equilibrium, which becomes a poorer and poorer approximation at increasing atmospheric heights.

The histogram displayed in Figure~\ref{figmhd2} illustrates that the majority of signs between the gradients calculated at iso-$z$ and iso-$\tau$ surfaces is preserved also in the degraded simulations, in particular at $\log \tau =0$. At this height about 16\% of the pixels have gradients that differ by angles between $0^{\circ}$ and $15^{\circ}$, being about 66\% of the pixels with the same sign. At $\log \tau =-1.5$, the pixels that differ by angles between $0^{\circ}$ and $15^{\circ}$ are about 11\%, with a total amount of the pixels having the same sign of about 57\%.

In conclusion, this analysis suggests that, at least in part, the observed variations of $P_{\mathrm {gas}}$ may have a physical origin.

\section{Discussion}

The evolution of continuum intensity maps of the EG shown in Figure~\ref{fig4} and in the online movie clearly indicates that we are observing a so-called \textit{active} granule, i.e., one which splits multiple times \citep[see][]{Oda:84}. The diameter reaches values of $4\arcsec - 5\arcsec$, in agreement with those found in previous studies.

Across this structure, magnetic flux emergence is taking place. Evidence of this is given by the appearance of adjacent mixed polarity patches expanding along the horizontal direction. Moreover, these patches generally have upward motions. The magnetic flux brought into the photosphere is of the order of $10^{18}$~Mx. The corresponding maximum value of the total unsigned flux ($2 \times 10^{18}$~Mx) places this solar feature at the frontier with the smallest ephemeral regions \citep[see][]{Lidia:15}.

Remarkably, we have hints of the existence of linear polarization signals in individual locations of the feature. In fact, the plots that are displayed in Figure~\ref{fig5} point out that Stokes~\textit{U} profiles well above the noise level are required in certain pixels to fit the observed quantity $V_{\mathrm{measured}}$, which is affected by a cross-talk with Stokes~$U_{\mathrm \sun}$. In addition, the distribution of significant Stokes~$U_{\mathrm{fitted}}$ over the EG (see Figure~\ref{figpol}) suggests that a large part of the structure contains transverse fields. This evidence was not observed by \citet{Palacios:12}, who did not take into account the cross-talk present in these IMaX L12-2 measurements. The only caveat is that some gradients of $v_{\mathrm LOS}$ and $B$ might have been interpreted in terms of transverse fields by the \textit{SirUV} code.

Concerning the thermodynamical properties of the EG, in Figures~\ref{fig7}, \ref{fig8}, and~\ref{fig8extra} we have presented plots of $T$, $P_{\mathrm {gas}}$, and $P_{\mathrm {mag}}$ derived from the atmospheric model retrieved by the \textit{SirUV} code. They suggest that in the observed EG the gas pressure is slightly smaller than in the very quiet Sun. 
More importantly, they make visible that the total pressure (gaseous + magnetic) increases by at least 5\% with respect to the very quiet Sun. Therefore, the magnetic field seems to provide an additional source for the horizontal expansion of the EG. 

However, we cannot exclude that a fraction of the field could be in the lanes surrounding the EG, actually stopping the granule from expanding. Moreover, we are able to evaluate the pressure at iso-$\tau$, rather than at iso-$z$ as it would be required to analyze magnetohydrostatic equilibrium in the structure. Thus, we cannot dismiss the possibility that the higher pressure at given $\tau$ could also arise because $\tau$ moves somewhat in height when there is a magnetic field. In our description, we also neglect dynamic pressure, which may play an important role in the evolution of large granules that may not be in hydrostatic equilibrium. Indeed, the horizontal flows in granules produce a force on their surrounding gas that acts similarly to pressure from one direction, which is larger for EGs than for ordinary granules \citep[see, e.g.][]{Ploner:99}. Lastly, note that our estimate of the magnetic pressure does not take into full account the contribution of $B_{\mathrm{tran}}$, which cannot be evaluated with this IMaX measurements, although it is expected to be significant and initially larger than $B_{\mathrm{long}}$ in the emerging flux region. 

For their part, numerical simulations predict that, at the sites of flux emergence, the additional magnetic pressure is able to sustain the horizontal expansion velocity of the granules, giving rise to abnormal granulation 
\citep{Cheung:07,Cheung:08,Sykora:08,Tortosa:09,Cheung:10,Nobrega:16}. Actually, abnormal granulation was found in high resolution observations of bipolar flux emergence 
(e.g., \citealp{Otsuji:07,Orozco:08,Guglielmino:10,Ortiz:14,Santiago:14,Guglielmino:18}; see also the review of \citealp*{Cheung:14}).

In connection with these findings, we can compare our results with those of \citet{Cheung:07} and \citet{Tortosa:09}. In particular, \citet{Cheung:07} used the MURaM code to carry out radiative MHD simulations of the emergence of magnetic flux tubes, using grey radiative transfer. In their simulation of a ``weak'' flux tube, with initial field strength at the tube axis of $2500$~G and total longitudinal flux of $3.1 \times 10^{18}$~Mx, they observed that the emergence of this magnetic flux tube does not lead to a severe disturbance in the appearance of the granulation. At the site of emergence, they found the existence of predominantly horizontal fields with strengths of up to $400$~G, having a rise velocity of $1-2$~\kms. Moreover, the morphology of the emerged field resembles a \textit{salt-pepper} pattern, displaying a mixture of positive and negative small-scale flux in the intergranular network. The flux contained in each individual polarity of the emerging flux concentration is about $8 \times 10^{17}$~Mx. This occurs because the flux tube is not sufficiently buoyant to rise coherently against the convective flows and fragments, leading to recirculation and overturning of material in the near-surface layers of the convection zone. 

Such characteristics are qualitatively in agreement with our IMaX observations, but in these simulations the granulation pattern seems to remain almost undisturbed. 

For their part, \citet{Tortosa:09} used the MURaM code to model the emergence of magnetized plasma across granular cells in the low solar atmosphere, including radiative transfer and a detailed equation of state. In their horizontal-tube experiment, with total longitudinal flux of $3.8 \times 10^{19}$~Mx, magnetized granules appear at the photosphere as isolated cells with upward velocity and grow in about 8 minutes reaching a maximum size of $7.5 \,\mathrm{Mm}^2$, much larger than non-magnetic granules. Upon arrival in the photosphere, the magnetic field is predominantly horizontal, with vertical footpoints in the integranular lanes. Then, the anomalous granules begin to fragment and two magnetic patches with vertical field of positive and negative polarity, respectively, appear in their interior. In the meantime, intergranules become more and more populated with vertical-field elements. Finally, fragmentation occurs like in normal granules, with the onset of downflow and the appearance of intergranular lanes that cut across the cells. The lifetime of the anomalous granulation is around 15 minutes. Moreover, in anomalous granular cells, the total pressure reaches a peak 20\% larger than in normal granules, due to the magnetic pressure present in the magnetized granules.  

All of the above mentioned values are compatible with our findings, except the order of magnitude difference in magnetic flux. Also the value of the horizontal expansion velocity found in these MHD simulations (about $4-6$~\kms), is slightly larger but still comparable with that obtained from our IMaX observations, of the order of 1~\kms. This estimate was reported by \citet{Palacios:12}, who calculated it for the same emergence event from geometrical considerations and from flowmaps of various quantities using a local correlation tracking method.

Recently, using high-resolution MHD simulations of a photospheric small-scale dynamo performed with the MURaM code, \citet{Rempel:18} analyzed the amplification of magnetic field in EGs. This study focused on the time evolution of newly formed downflow lanes in the centers of EGs, finding that horizontal flows converging toward the lanes can amplify an initial weak vertical field of about a few $10$~G up to $800$~G within a few minutes. This field is organized on extended, narrow magnetic sheets having a length comparable to the granular scale. This process appears to be due to the contributions of both shallow and deep recirculation. In particular, the deep recirculation seems to be linked to strong sheet-like structures, coupling the magnetic field that reaches the photosphere in the centers of EGs to the deeper convection zone, while the shallow recirculation provides the primary source for the subsequent appearance of small-scale turbulent fields in the downflow lanes.

\begin{figure}[!t]
	\centering
	\includegraphics[trim= 20 550 280 235, clip, scale=.57]{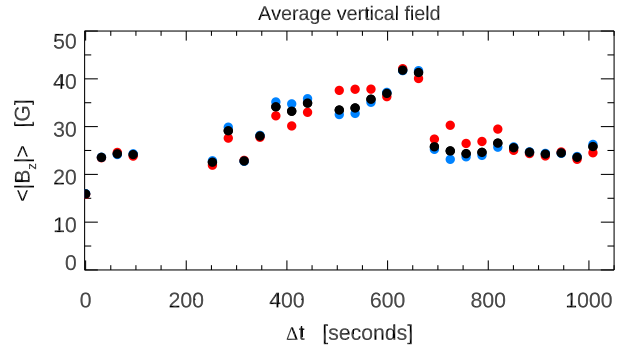}
	\caption{Evolution of the average absolute value of the longitudinal component of the magnetic field in the emerging structure cospatial to the EG (black symbols). Blue and red symbols refer to the average values in regions with upflows or downflows, respectively. \label{figrempel}}
\end{figure}

In the numerical simulations of \citet{Rempel:18}, upflow regions are dominated by a horizontal field. Allowing deep recirculation, the average value for $|B_{\mathrm{z}}|$ and $|B_{\mathrm{h}}|$ at $\log \tau =0$ in these upflow regions are $50$~G and $100$~G, respectively. Field amplification in the downflow lanes leads to average values of $130$~G and $160$~G for $|B_{\mathrm{z}}|$ and $|B_{\mathrm{h}}|$, respectively. In an attempt to compare our observational results with these numerical findings, we have calculated the average value of $|B_{\mathrm{z}}| \equiv |B_{\mathrm{long}}|$ at $\log \tau =0$ within the EG. Figure~\ref{figrempel} shows the trend of $\langle\left|B_{\mathrm{z}}\right|\rangle$ for the entire area of the EG (black symbols) and for the upflow/downflow regions (blue/red symbols). We find that $\langle\left|B_{\mathrm{z}}\right|\rangle$ is stronger in the upflow regions during the first phases of the analyzed EG, whereas at later times $\langle\left|B_{\mathrm{z}}\right|\rangle$ becomes stronger in the downflow regions. We also note a resemblance between the numerical results and our observations in the shape of the magnetic sheet cospatial to the central downflow lanes observed at $\Delta t = 504$~s in Figure~\ref{fig4}. Nonetheless, one has to be cautious when comparing the numerical values from the study of \citet{Rempel:18} with the observed values, which appear to fall short of the former. Indeed, it would be necessary to degrade the results of simulations to mimic the spatial resolution of the IMaX observations. Moreover, due to the lack of full spectropolarimetric measurements we are not able to estimate the value of the horizontal field $|B_{\mathrm{h}}|$, which is expected to be dominant in the EG.


In this context, it is important to mention that, recently, \citet{Moreno:18} identified the formation of organized horizontal magnetic sheets covering whole granules in realistic 3D magnetoconvection models of small-scale flux emergence. In this regard, \citet{Fischer:19} reported on the first clear observational evidence of a magnetic sheet's emergence, characterizing its development. \citet{Moreno:18} also provided a rough estimation for the occurrence of sheet-like events, between 0.3 and 1~$\mathrm{day}^{-1}\,\mathrm{Mm}^{-2}$. Given the IMaX FoV (about $46\arcsec \times 46\arcsec$) and the duration of the time series ($\approx 20$~minutes of stable pointing), with an intermediate frequency value of 0.5~$\mathrm{day}^{-1}\,\mathrm{Mm}^{-2}$ we would expect to observe about 8 events. However, we only found this single event in the IMaX L12-2 time series. 

Emergence events with horizontal, i.e., very inclined magnetic fields, such as the one here investigated looks like, might be somehow related with HIFs, as already suggested by \citet{DePontieu:02}. Indeed, \citet{Lites:08} reported on HIFs organized on a mesogranular scale. The presence of such a cellular pattern on a scale of $5\arcsec-10\arcsec$ was confirmed by \citet{Ishikawa:10}, who further linked the transient horizontal fields to the emergence of granular-scale horizontal fields, advected by mesogranular flows \citep[see also][]{Ishikawa:11}. Mesogranular boundaries were found to be the preferential location of magnetic elements \citep{Lotfi:11}, in particular at the site of convectively driven sinks \citep{Requerey:17}, as also reported in  simulations of small-scale dynamo action in the quiet Sun \citep[e.g.,][]{Bushby:14}. In addition, cluster emergence of mixed polarities in the internetwork, carrying a flux amount of about $10^{18}$~Mx, was observed by \citet{Wang:12}.

The origin of these magnetic structures may reside in weakly twisted magnetic fields, which do not survive into the photosphere as coherent flux bundles. This is shown in numerical models of flux tubes rising from the bottom of the convection zone \citep{Moreno:95,Magara:01,Toriumi:10,Sykora:15}. The remnant fields can be recycled by magnetoconvection to form internetwork fields. Alternatively, the remnants of decaying active regions can be also recycled, as suggested by \citet{Ploner:01}. Nonetheless, this is surely not the case in these IMaX observations, which were taken during the long solar minimum in 2008-2010. As a third option, the generation of such emergence events with horizontal fields in the quiet Sun may be ascribed to a turbulent local dynamo process occurring in the near-surface layers \citep[see, e.g.,][and references therein]{Vogler:07,Rempel:14,Borrero:15}. In this case, possibly such small-scale magnetic fields would not have a relationship with the global dynamo, albeit recent dynamo simulations indicate a tangling between a small-scale dynamo mechanism and the large-scale field \citep{Karak:16}.


\section{Conclusions}

IMaX/Sunrise observations of the solar photosphere taken at disk center have revealed a number of small-scale episodes of magnetic flux emergence \citep[see, e.g.,][and references therein]{Danilovic:10,Solanki:10,Solanki:17,Guglielmino:12}. 

Here, we have studied the emergence phase of a magnetic structure, extending over mesogranular scale, being cospatial with an EG. We have analyzed the polarization maps and, then, we have inverted the Stokes profiles with the SIR code, to obtain information on the physical parameters of the magnetic structure. Our study reveals that this feature hosts linear polarization patches and brings into the photosphere an excess of total pressure (gaseous + magnetic). 

The overall characteristics suggest that we are observing an emerging multi-polar \textit{magnetic flux-sheet} structure, which seems to be able to disturb the granulation pattern. The detection of this feature is made possible by the high polarimetric sensitivity of IMaX.

Our findings are in agreement with numerical simulations of flux emergence at small scale, some of which indicate the presence of anomalous granular cells in these sites. This could also be the case for some EGs, like the one here analyzed: if the incipient granule carries a magnetic field of sufficient intensity, the additional push of the magnetic pressure might lead to the formation of cells larger than usual. A magnetic field strength of the order of the equipartion value with the kinetic energy of photospheric plasma flows, $\approx 400$~G \citep[see, e.g.,][]{Solanki:96,Ishikawa:08}, seems to be necessary to see these disturbances in the granulation \citep{Cheung:07}. 

Indeed, state-of-the-art simulations do not clearly show if the amount of flux carried by the structure that we observe is enough to modify the granulation pattern. The emergence event studied with our IMaX observations appears to have characteristics intermediate between those found in the MHD numerical experiments of \citet{Cheung:07} and those of \citet{Tortosa:09}. This points out that simulations of emergence of such amounts of flux are needed to properly interpret solar features like the present one. In addition, note that recent numerical models of flux emergence including magnetoconvection have shown the existence of small-scale magnetic sheets that at maximum development cover granular surfaces \citep{Moreno:18}. 

However, we cannot rule out the possibility that we are partly also seeing weak fields that were present earlier \citep{Lamb:08,Lamb:10}, as discussed in \citet{Gosic:16}. These could become concentrated by the shuffling motions due to the EG and, hence, visible. For instance, this seems to occur in the EG studied by \citet{Fischer}. Moreover, we should also consider the possibility that the presence of the EG, which may have a large upflow under it, is helping the magnetic flux to emerge at its location, rather than the emerging field driving the evolution of the EG.  

In order to verify whether EGs have a magnetic nature, it would be worthwhile to study the frequency of flux emergence events cospatial with EGs, to determine how often these occur. Such an endeavour could start by checking if other EGs present in IMaX observations, taken during both the first and the second \textsc{Sunrise} flights, are associated with emerging flux. More in general, full spectropolarimetric measurements with high spectral resolution would be a more straightforward approach to identify the presence of horizontal fields associated with EGs. 
The high spatial resolution, high polarimetric sensitivity, and long temporal coverage provided by the PHI instrument \citep{Solanki:15,Solanki:20} on board the Solar Orbiter space mission \citep{Muller:13} will surely enhance our capability to find these events and increase the statistics for such a study. To this purpose, it will be mandatory to use an effective segmentation algorithm for pattern recognition to identify EGs in solar granulation and compare their location with horizontal flux emergence events. In this context, a method used to detect families of splitting granules like the one developed by \citet{Roudier:03,Roudier:16} to study families of fragmenting granules appears to be more promising than classical approaches based on edge or intensity levels such those described, e.g., by \citet{Berrilli:05} and \citet{Falco:17}. Finally, multi-line measurements would also allow analyzing the thermodynamical properties in more detail at different atmospheric heights. 

New investigations on the properties of these structures are required to find an answer about their origin and to reach a unified understanding of their dynamics. In particular, the analysis of the frequency of these emergence events in correlation with the solar cycle will provide a clearer picture of their sources. This will be achieved by using data from the PHI instrument as well as from the large-aperture ground-based telescopes DKIST \citep{Keil:10} and EST \citep{Collados:10}.

\acknowledgments

This research has received funding from the European Commission's Seventh Framework Programme under the grant agreement no.~312495 (SOLARNET project) and from the European Union's Horizon 2020 research and innovation programme under the grant agreements no.~739500 (PRE-EST project) and no.~824135 (SOLARNET project). S.L.G. and F.Z. acknowledge support by the Italian MIUR-PRIN grant 2017APKP7T on \textit{Circumterrestrial Environment: Impact of Sun-Earth Interaction}, by the Universit\`{a} degli Studi di Catania (Piano per la Ricerca Universit\`{a} di Catania 2016-2018 -- Linea di intervento~1 ``Chance''; Linea di intervento~2 ``Dotazione ordinaria''; Fondi di Ateneo 2020-2022, Universit\`{a} di Catania, Linea Open Access), by the Istituto Nazionale di Astrofisica (PRIN INAF 2010/2014), and by Space Weather Italian COmmunity (SWICO) Research Program. 
This project has received funding from the European Research Council (ERC) under the European Union’s Horizon 2020 research and innovation programme (grant agreement no.~695075) and has been supported by the BK21 plus program through the National Research Foundation (NRF) funded by the Ministry of Education of Korea. 
The National Solar Observatory (NSO) is operated by the Association of Universities for Research in Astronomy, Inc.~(AURA), under cooperative agreement with the National Science Foundation.
This work has been partially funded by the Spanish Ministerio de Educaci\'on y Ciencia, through Projects ESP2006-13030-C06-01/02/03/04 and AYA2009-14105-C06, by the Spanish Ministry of Economy and Competitiveness through Project ESP-2016-77548-C5, and by Junta de Andaluc\'ia, through Project P07-TEP-2687, including a percentage from European FEDER funds. 
L.R.B.R. and J.C.T.I. acknowledge financial support from the State Agency for Research of the Spanish Ministerio de Ciencia, Innovaci\'on y Universidades through the ``Center of Excellence Severo Ochoa'' award to the Instituto de Astrof\'isica de Andaluc\'ia (SEV-2017-0709).
The German contribution to \textsc{Sunrise} is funded by the Bundesministerium f\"ur Wirtschaft und Technologie through Deutsches Zentrum f\"ur Luft- und Raumfahrt e.V.~(DLR), Grant no.~50 OU 0401, and by the Innovationsfond of the President of the Max Planck Society (MPG). 
Use of NASA's Astrophysical Data System is gratefully acknowledged.
This article honors the memory of Prof.~Egidio Landi degl'Innocenti, who passed away while this paper was being written, the most important results being discussed during the Solar Polarization Workshop 8 held in his honor.

\facility{\textsc{Sunrise}}.

\appendix

We have carried out a series of preliminary runs to test the reliability of the \emph{SirUV} inversions. We have taken 2000 atmosphere models, randomly chosen from those obtained from the inversions of an IMaX V5-6 data set \citep[see][]{Guglielmino:12}. To avoid the tendency of the inversions that retrieve horizontal fields for pixels at noise level \citep[see, e.g.][]{Asensio:09,Stenflo:10,Basilio:16}, we have imposed the condition that the original inclination had to be $\gamma < 80^{\circ}$ or $\gamma > 100^{\circ}$ in order for the model to be included in the set. These models have been used as input for the synthesis of 2000 profiles of Stokes~\emph{I}, \emph{U},~and~\emph{V}. The range used for the stratification is $-4.0 < \log \tau < 1.4$, where $\tau$ is the optical depth at $500$~nm. The computed synthetic profiles have been convolved with the spectral Point Spread Function at the focal plane of IMaX. Moreover, we have added random noise to these synthetic profiles, which corresponds to the signal-to-noise ratio for these IMaX V5-6 measurements, about $1 \times 10^{-3}$ in units of $I_c$ per wavelength point in each Stokes parameter. 

\begin{figure*}[b]
	\centering
	\includegraphics[trim=0 425 0 0, clip, scale=.85]{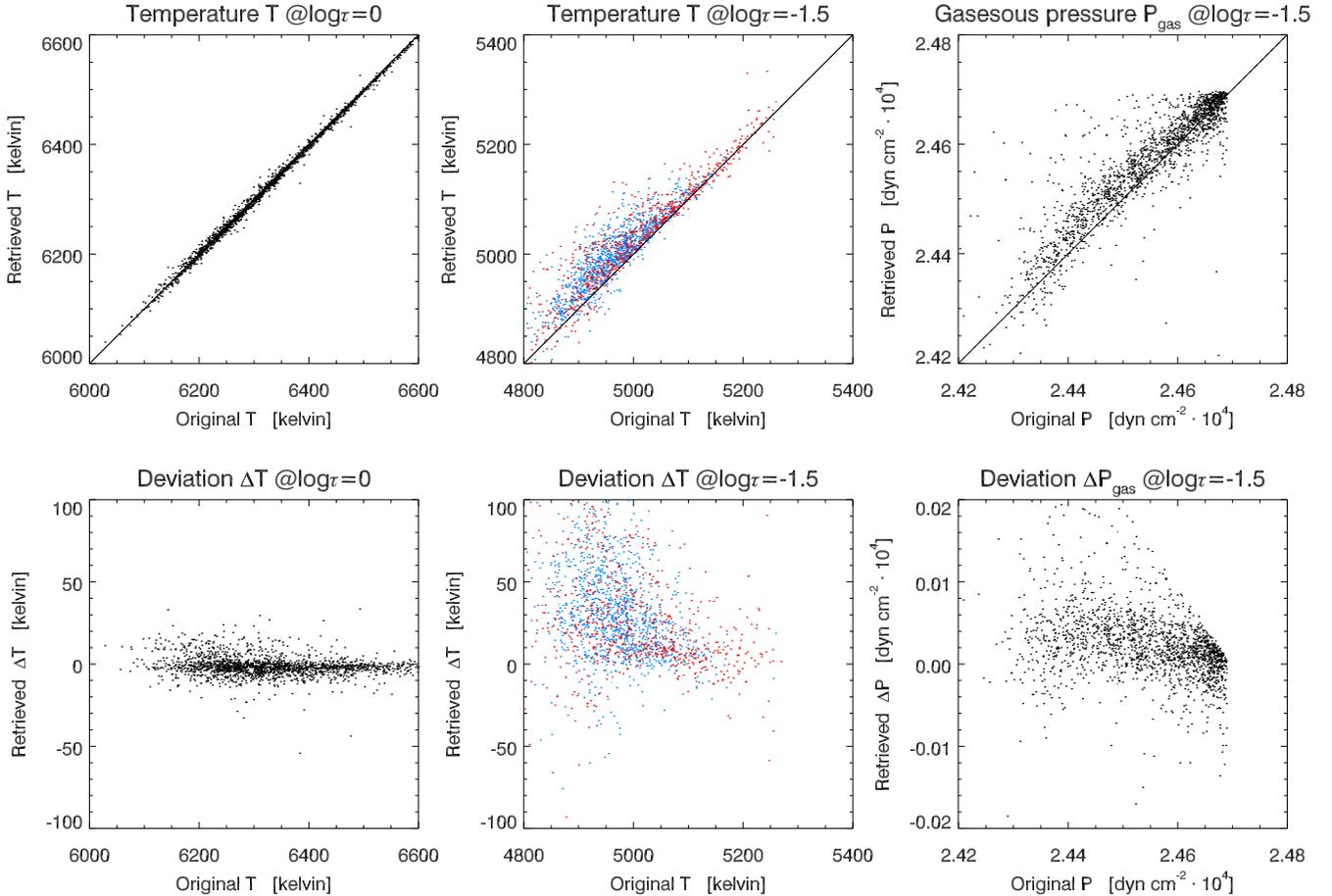}
	\caption{\textit{Top}: Scatter plots of the temperature retrieved by the \emph{SirUV} code vs. the original atmospheric value (see main text for more details), at $\log \tau=0$ (left) and $\log \tau=-1.5$ (middle). Same, for the gas pressure at $\log \tau=-1.5$ (right). \textit{Bottom}: The deviations between the original atmospheric value and the values retrieved by the inversions, for the same quantities above mentioned. Blue (red) points refer to original atmospheric models with $B < 100$~G ($B > 100$~G). \label{fig1}}
\end{figure*}

\begin{figure*}[!t]
	\centering
	\includegraphics[trim=0 175 0 5, clip, scale=.85]{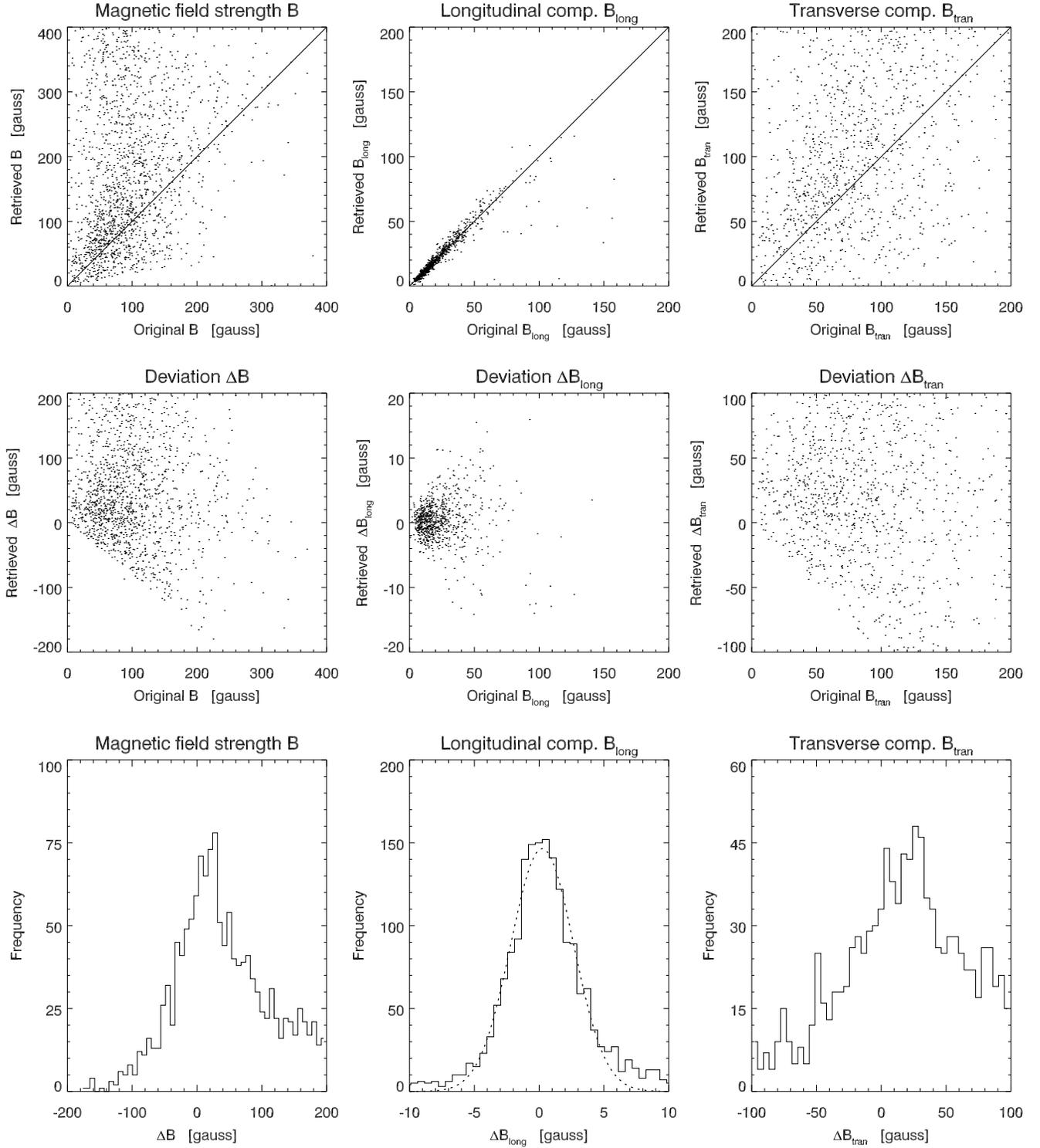}
	\caption{\textit{Top}: Scatter plots of $B$ (left), $B_{\mathrm{long}}$ (middle), and $B_{\mathrm{tran}}$ (right) retrieved by the \emph{SirUV} code vs. the original atmospheric values (see main text for more details). \textit{Middle}: The deviations between the original atmospheric value and the values retrieved by the inversions, for the above mentioned quantities. \textit{Bottom}: The corresponding frequency histograms, for the differences plotted in the middle row. The dashed line overplotted on the histogram of $B_{\mathrm{long}}$ represents a Gaussian fit to the data. \label{fig2}}
\end{figure*}

\begin{figure*}[!t]
	\centering
	\includegraphics[trim=0 175 0 5, clip, scale=.85]{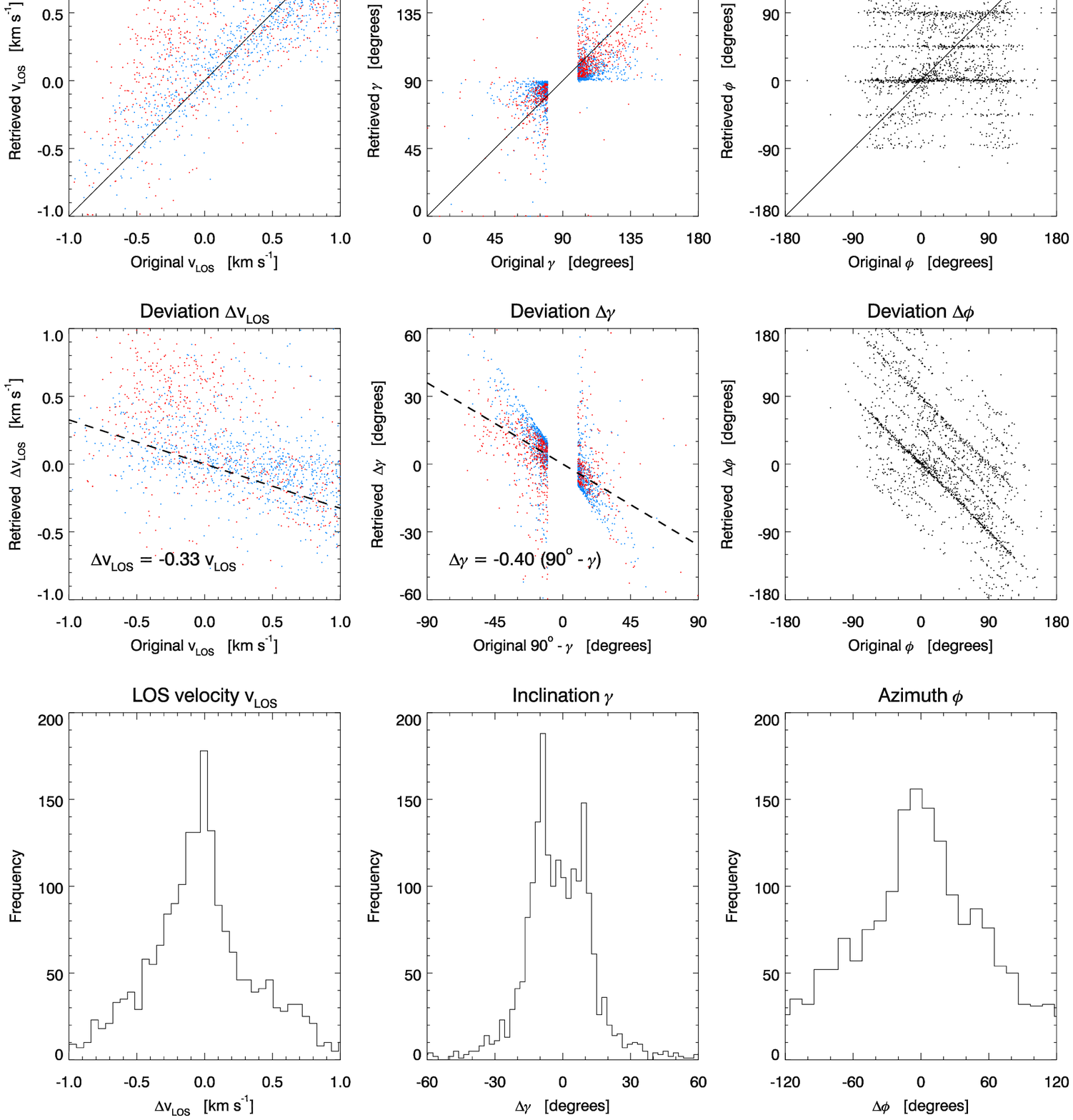}
	\caption{Same as in Figure~\ref{fig2}, for $v_{\mathrm{LOS}}$, $\gamma$, and $\phi$. Blue (red) points refer to original atmospheric models with $B < 100$~G ($B > 100$~G). Dashed lines overplotted on the deviation of $v_{\mathrm{LOS}}$ and $\gamma$ represents a linear fit to the data. \label{fig3}}
\end{figure*}

As the next step, we have generated the linear combination represented by Equation~(\ref{eqUV}) using the synthetic Stokes~\emph{U} and~\emph{V} parameters. This computed profile and the Stokes~\emph{I} profile have been sampled at the same wavelength positions used in the L12-2 observing mode. Then, we have provided the \emph{SirUV} code 
with these simulated data, inverting the Stokes~\emph{I} profile and the linear combination in Equation~(\ref{eqUV}). The Harvard Smithsonian Reference Atmosphere \citep[HSRA;][]{Gingerich:71} has been used as initial guess for the temperature stratification. For the inversion, we have used 3 iteration cycles, with up to 4 nodes for the temperature $T$, 2 nodes for the LOS velocity $v_{\mathrm LOS}$ and $B$, and 1 node for the other parameters. No filling factor is taken into account in the inversion. The total number of free parameters is nine for the \emph{SirUV} inversions, with a total number of 24 data points to be considered.

Finally, for each simulated and inverted profile, we have compared the original atmospheric values and those retrieved by the \emph{SirUV} inversion. We have also computed the deviation $\Delta$ between these values for several atmospheric parameters: $T$, electronic pressure, 
gas pressure $P_{\mathrm{gas}}$, microturbulent velocity, 
$v_{\mathrm LOS}$, $B$, $\gamma$, and azimuth angle $\phi$. The values of $v_{\mathrm LOS}$ and $B$ have been averaged between $\log \tau = -1$ and $\log \tau = -2$, where the response functions are more sensitive. 
We have further examined the longitudinal and transverse components of the magnetic field, $B\,\cos\gamma$ (= $B_{\mathrm{long}}$) and $B\,\sin\gamma$ (= $B_{\mathrm{tran}}$), respectively.

Figure~\ref{fig1} (top panels) displays the scatter plots of the temperature vs.~the original value at two different values of $\log \tau$, namely $\log \tau = 0$ (left) and $\log \tau = -1.5$ (middle), respectively. We find a very good agreement between the temperature derived by the \emph{SirUV} code and the original one at $\log \tau = 0$. At this height, their deviation $\Delta T$ is of the order of a few tens of kelvin (see Figure~\ref{fig1}, bottom left panel). The agreement is satisfactory also at $\log \tau = -1.5$. In this case, we notice a modest difference between the two populations: profiles emerging from atmospheric models with originally $B < 100$~G (blue points) present a slightly larger positive deviation than profiles with originally $B > 100$~G (red points). However, $\Delta T$ is not larger than 100 K and can be neglected (Figure~\ref{fig1}, bottom central panel). The scatter plot of $P_{\mathrm{gas}}$ at $\log \tau = -1.5$ is also shown (Figure~\ref{fig1}, top right panel). There is a pretty fair agreement between the original and inverted values of $P_{\mathrm{gas}}$ as well. For this parameter, the deviations are of the order of $0.5\%$ (Figure~\ref{fig1}, bottom right panel). Similar results are obtained for the electronic pressure and the microturbulent velocity. 

In Figures~\ref{fig2}~and~\ref{fig3} we show the scatter plots for other physical quantities, their deviations, as well as the frequency histograms of the deviations. 

Figure~\ref{fig2} presents the graphs for the total magnetic field strength $B$ (first column) and for the components $B\,\cos\gamma$ (second column) and $B\,\sin\gamma$ (third column). The magnetic field strength plots indicate considerable scatter and a tendency of the \emph{SirUV} inversions to overestimate the value of $B$. This may depend on the noise in the data \citep{Borrero:11,Borrero:12}. The longitudinal component $B_{\mathrm{long}}$ exhibits an excellent agreement between the original and inverted values. The deviation has a symmetric, very narrow Gaussian profile, as can be seen from Figure~\ref{fig2} (bottom middle panel). The full-width at half maximum (FWHM) is about $5.7$~G. In contrast, the transverse component shows a very wide scatter. This leaves $B_{\mathrm{tran}}$ totally undetermined. Note that, for the magnetic field and its components, there is no dependence on the values of the original values of $B$. The large uncertainty in $B_{\mathrm{tran}}$ is not surprising, given that the full Stokes vector is not measured, further compounded by the relatively small number of sampled wavelength points. That is also the main reason for the large uncertainty in determining $B$.

Figure~\ref{fig3} displays the plots for $v_{\mathrm{LOS}}$ (first column), $\gamma$ (second column), and $\phi$ (third column). For $v_{\mathrm {LOS}}$ and $\gamma$, we again find the difference in the scatter plots between the inversions of profiles with original $B < 100$~G (blue points) and those with $B > 100$~G (red points). The LOS velocity is slightly better retrieved for profiles with weaker original field strength. For profiles with $B < 100$~G, we find a linear relation between $v_{\mathrm{LOS}}$ and $\Delta v_{\mathrm {LOS}}$

\[
\Delta v_{\mathrm {LOS}} = -0.33 \,\, v_{\mathrm{LOS}}
\]

\noindent
with an estimated error on the coefficient of $\pm 0.01$. The histogram of the cumulative deviations for $v_{\mathrm{LOS}}$ has a roughly symmetric profile (Figure~\ref{fig3}, bottom left panel). The best-fitting Gaussian distribution has a FWHM of about $0.74$~\kms. As far as the inclination is concerned, we note a general tendency of the inversions to retrieve a value of $\gamma$ that is more horizontal than the original one. This behaviour is reflected in the double peaked histogram (Figure~\ref{fig3}, bottom middle panel), which is likely an artifact produced by imposing the condition on the original inclination to be $\gamma < 80^{\circ}$ or $\gamma > 100^{\circ}$, and in the scatter plot of the deviation with respect to the original value of the quantity $(90^{\circ} - \gamma)$ (Figure~\ref{fig3}, central middle panel). The tendency is stronger for weaker field profiles, as commonly found in inversions of noisy polarization signals \citep{Borrero:11,Borrero:12}. For profiles with $B > 100$~G, we are able to infer a linear relation between $\gamma$ and $\Delta\gamma$, that is given by

\[
\Delta\gamma = -0.40 \,\, (90^{\circ}-\gamma)
\]

\noindent
with an estimated error on the coefficient of $\pm 0.03$. For the sake of precision, it has to be pointed out that this formula might be affected by the restriction to original inclination $\gamma < 80^{\circ}$ or $\gamma > 100^{\circ}$. Lastly, we note that also the azimuth angle remains completely indeterminate. 

This preliminary test allows us to argue that the \emph{SirUV} code is able to give reliable values only for certain atmospheric parameters from the inversion of Stokes $I$ and of the linear combination of Stokes \emph{U} and \emph{V} described in Equation~(\ref{eqUV}). First, the code infers the correct values of the thermodynamic parameters. Indeed, such a behaviour of the code is expected, because these physical quantities leave their mark on the Stokes~\emph{I} profile at least as strongly as on the other Stokes parameters. Furthermore, the \emph{SirUV} code provides an excellent estimate of the longitudinal component of the magnetic field, $B\,\cos\gamma$. 

Somewhat surprisingly, we obtain a poor estimate for $v_{\mathrm{LOS}}$, which is better recovered for profiles with weaker original field strength. This fact reflects that, in these profiles, the LOS velocity is mainly inferred from the Stokes~\emph{I} parameter, with limited disturbance due to the signal of Stokes~$U_{\mathrm \sun}$ which alters the Stokes~\emph{V} signal. 
Also, the code gives inadequate information on the values of $B$ and $\gamma$. The former is often overestimated in comparison to the original value, the latter is more horizontal in most cases, in particular in the profiles with weaker original field that are closer to the noise level and have weaker Stokes~\emph{V} signals. Finally, the transverse component of the magnetic field and $\phi$ remain completely undetermined. 

The inability of the \emph{SirUV} inversions to recover the correct values for these quantities can be easily understood. $\gamma$ is fundamentally given by the ratio between the linear polarization ($\sqrt{Q^2+U^2}$) and the circular polarization ($V$). As well, $\phi$ depends on the ratio between $Q$ and $U$. Provided that the code is not supplied with independent information on the linear and circular polarization, in particular it totally lacks of knowledge of Stokes~\emph{Q}, it cannot estimate these physical parameters in a reliable manner. Conversely, the longitudinal component of the magnetic field is basically dependent on $V$, which is entirely supplied to the code via the linear combination described in Equation~(\ref{eqUV}).

In conclusion, the \emph{SirUV} code wipes away the effect of the residual polarization cross-talk in $V_{\mathrm{measured}}$ as far as the measurements of the longitudinal flux is concerned. Thus, the magnetic flux density can be correctly evaluated. 
Note that we have carried out these tests by choosing only atmospheric models with original weak field strength, $B < 500$~G, as we have intended to use the \emph{SirUV} code in analogous conditions over quiet-Sun internetwork regions observed by IMaX in our dataset. 


\begin{thebibliography}{}


\bibitem[Allen \& Musman(1973)]{Allen:73} Allen, M.~S., \& Musman, S.\ 1973, \solphys, 32, 311 

\bibitem[Anusha et al.(2017)]{Anusha:17} Anusha, L.~S., Solanki, S.~K., Hirzberger, J., \& Feller, A.\ 2017, \aap, 598, A47 

\bibitem[Asensio Ramos(2009)]{Asensio:09} Asensio Ramos, A.\ 2009, \apj, 701, 1032 


\bibitem[Barthol et al.(2011)]{Barthol:11} Barthol, P., Gandorfer, A., Solanki, S.~K., et al.\ 2011, \solphys, 268, 1 


\bibitem[Berkefeld et al.(2011)]{Berkefeld:11} Berkefeld, T., Schmidt, W., Soltau, D., et al.\ 2011, \solphys, 268, 103 

\bibitem[Berrilli et al.(2002)]{Berrilli:02} Berrilli, F., Consolini, G., Pietropaolo, E., et al.\ 2002, \aap, 381, 253 

\bibitem[Berrilli et al.(2005)]{Berrilli:05} Berrilli, F., Del Moro, D., Russo, S., et al.\ 2005, \apj, 632, 677

\bibitem[Borrero et al.(2015)]{Borrero:15} Borrero, J.~M., Jafarzadeh, S., Sch{\"u}ssler, M., \& Solanki, S.~K.\ 2015, \ssr,  

\bibitem[Borrero \& Kobel(2012)]{Borrero:12} Borrero, J.~M., \& Kobel, P.\ 2012, \aap, 547, A89 

\bibitem[Borrero \& Kobel(2011)]{Borrero:11} Borrero, J.~M., \& Kobel, P.\ 2011, \aap, 527, A29 

\bibitem[Bushby \& Favier(2014)]{Bushby:14} Bushby, P.~J., \& Favier, B.\ 2014, \aap, 562, A72

\bibitem[Carlier et al.(1968)]{Carlier:68} Carlier, A., Chauveau, F.,
Hugon, M., R\"osch, J.\ 1968, Academie des Sciences Paris Comptes Rendus Serie B Sciences Physiques, 266, 199 

\bibitem[Cheung \& Isobe(2014)]{Cheung:14} Cheung, M.~C.~M., \& Isobe, H.\ 2014, Living Reviews in Solar Physics, 11, 3

\bibitem[Cheung et al.(2010)]{Cheung:10} Cheung, M.~C.~M., Rempel, M., Title, A.~M., \& Sch\"ussler, M.\ 2010, \apj, 720, 233

\bibitem[Cheung et al.(2008)]{Cheung:08} Cheung, M.~C.~M., Sch{\"u}ssler, M., Tarbell, T.~D., \& Title, A.~M.\ 2008, \apj, 687, 1373

\bibitem[Cheung et al.(2007)]{Cheung:07} Cheung, M.~C.~M., Sch{\"u}ssler, M., \& Moreno-Insertis, F.\ 2007, \aap, 467, 703


\bibitem[Collados et al.(2010)]{Collados:10} Collados, M., Bettonvil, F., Cavaller, L., \& EST Team 2010, AN, 331, 615

\bibitem[Danilovic et al.(2010)]{Danilovic:10} Danilovic, S., Beeck, B., Pietarila, A., et al.\ 2010, \apjl, 723, L149 


\bibitem[del Toro Iniesta \& Ruiz Cobo(2016)]{Basilio:16} del Toro Iniesta, J.~C., \& Ruiz Cobo, B.\ 2016, Living Reviews in Solar Physics, 13, 4 

\bibitem[De Pontieu(2002)]{DePontieu:02} De Pontieu, B.\ 2002, \apj, 569, 474 

\bibitem[Dom\'inguez Cerde\~na(2003)]{Domi:03} Dom\'inguez Cerde\~na, I.\ 2003, \aap, 412, L65 

\bibitem[Dravins et al.(1981)]{Dravins:81} Dravins, D., Lindegren, L., \& Nordlund, A.\ 1981, \aap, 96, 345 

\bibitem[Falco et al.(2017)]{Falco:17} Falco, M., Puglisi, G., Guglielmino, S.~L., et al.\ 2017, \aap, 605, A87

\bibitem[Fischer et al.(2019)]{Fischer:19} Fischer, C.~E., Borrero, J.~M., Bello Gonz{\'a}lez, N., et al.\ 2019, \aap, 622, L12

\bibitem[Fischer et al.(2017)]{Fischer} Fischer, C.~E., Bello Gonz{\'a}lez, N., \& Rezaei, R.\ 2017, \aap, 602, L12 

\bibitem[Gadun et al.(2000)]{Gadun:00} Gadun, A.~S., Hanslmeier, A., Pikalov, K.~N., et al.\ 2000, \aaps, 146, 267

\bibitem[Gandorfer et al.(2011)]{Gandorfer:11} Gandorfer, A., Grauf, B., Barthol, P., et al.\ 2011, \solphys, 268, 35 

\bibitem[Gingerich et al.(1971)]{Gingerich:71} Gingerich, O., Noyes, R.~W., Kalkofen, W., \& Cuny, Y.\ 1971, \solphys, 18, 347

\bibitem[Go{\v s}i{\'c} et al.(2016)]{Gosic:16} Go{\v s}i{\'c}, M., Bellot Rubio, L.~R., del Toro Iniesta, J.~C., Orozco Su{\'a}rez, D., \& Katsukawa, Y.\ 2016, \apj, 820, 35


\bibitem[Grossmann-Doerth et al.(1988)]{Schussler:88} Grossmann-Doerth, U., Sch\"ussler, M., \& Solanki, S.~K.\ 1988, \aap, 206, L37

\bibitem[Guglielmino et al.(2018)]{Guglielmino:18} Guglielmino, S.~L., Zuccarello, F., Young, P.~R., Murabito, M. \& Romano, P.\ 2018, \apj, 856, 127

\bibitem[Guglielmino et al.(2012)]{Guglielmino:12} Guglielmino, S.~L., Mart{\'{\i}}nez Pillet, V., Bonet, J.~A., et al.\ 2012, \apj, 745, 160 

\bibitem[Guglielmino et al.(2010)]{Guglielmino:10} Guglielmino, S.~L., Bellot Rubio, L.~R., Zuccarello, F., et al.\ 2010, \apj, 724, 1083

\bibitem[Hirzberger(2002)]{Hirz:02} Hirzberger, J.\ 2002, \aap, 392, 1105. 

\bibitem[Hirzberger et al.(2001)]{Hirz:01} Hirzberger, J., Koschinsky, M., Kneer, F., \& Ritter, C.\ 2001, \aap, 367, 1011 

\bibitem[Hirzberger et al.(1999a)]{Hirz:99a} Hirzberger, J., Bonet, J.~A., V\'azquez, M., \& Hanslmeier, A.\ 1999, \apj, 515, 441

\bibitem[Hirzberger et al.(1999b)]{Hirz:99b} Hirzberger, J., Bonet, J.~A., V\'azquez, M., \& Hanslmeier, A.\ 1999, \apj, 527, 405 

\bibitem[Hirzberger et al.(1997)]{Hirz:97} Hirzberger, J., V\'azquez, M., Bonet, J.~A., Hanslmeier, A., \& Sobotka, M.\ 1997, \apj, 480, 406 

\bibitem[Illing et al.(1975)]{Illing:75} Illing, R.~M.~E., Landman, D.~A., \& Mickey, D.~L.\ 1975, \aap, 41, 183

\bibitem[Ishikawa \& Tsuneta(2011)]{Ishikawa:11} Ishikawa, R., \& Tsuneta, S.\ 2011, \apj, 735, 74

\bibitem[Ishikawa \& Tsuneta(2010)]{Ishikawa:10} Ishikawa, R., \& Tsuneta, S.\ 2010, \apjl, 718, L171 


\bibitem[Ishikawa et al.(2008)]{Ishikawa:08} Ishikawa, R., Tsuneta, S., Ichimoto, K., et al.\ 2008, \aap, 481, L25 

\bibitem[Jafarzadeh et al.(2014)]{Jafarzadeh:14} Jafarzadeh, S., Cameron, R.~H., Solanki, S.~K., et al.\ 2014, \aap, 563, A101

\bibitem[Jafarzadeh et al.(2013)]{Jafarzadeh:13} Jafarzadeh, S., Solanki, S.~K., Feller, A., et al.\ 2013, \aap, 549, A116 




\bibitem[Karak \& Brandenburg(2016)]{Karak:16} Karak, B.~B., \& Brandenburg, A.\ 2016, \apj, 816, 28 

\bibitem[Kawaguchi(1980)]{Kawaguchi:80} Kawaguchi, I.\ 1980, \solphys, 65, 207 

\bibitem[Keil et al.(2010)]{Keil:10} Keil, S. L., Rimmele, T. R., Wagner, J., \& ATST Team 2010, AN, 331, 609

\bibitem[Kitai \& Kawaguchi(1979)]{Kitai:79} Kitai, R., \& Kawaguchi, I.\ 1979, \solphys, 64, 3 

\bibitem[Lagg et al.(2010)]{Lagg:10} Lagg, A., Solanki, S.~K., Riethm{\"u}ller, T.~L., et al.\ 2010, \apjl, 723, L164 

\bibitem[Lamb et al.(2010)]{Lamb:10} Lamb, D.~A., DeForest, C.~E., Hagenaar, H.~J., Parnell, C.~E., \& Welsch, B.~T.\ 2010, \apj, 720, 1405 

\bibitem[Lamb et al.(2008)]{Lamb:08} Lamb, D.~A., DeForest, C.~E., Hagenaar, H.~J., Parnell, C.~E., \& Welsch, B.~T.\ 2008, \apj, 674, 520

\bibitem[Landi Degl'Innocenti(1992)]{Egidio} Landi Degl'Innocenti, E.\ 1992, Solar Observations: Techniques and Interpretation, 71 

\bibitem[Lites et al.(2008)]{Lites:08} Lites, B.~W., Kubo, M., Socas-Navarro, H., et al.\ 2008, \apj, 672, 1237

\bibitem[Lites et al.(1996)]{Lites:96} Lites, B.~W., Leka, K.~D., Skumanich, A., Mart{\'{\i}}nez Pillet, V., \& Shimizu, T.\ 1996, \apj, 460, 1019 

\bibitem[Magara(2001)]{Magara:01} Magara, T.\ 2001, \apj, 549, 608 

\bibitem[Mart{\'{\i}}nez Gonz{\'a}lez et al.(2012a)]{Marian:12} Mart{\'{\i}}nez Gonz{\'a}lez, M.~J., Bellot Rubio, L.~R., Solanki, S.~K., et al.\ 2012, \apjl, 758, L40 

\bibitem[Mart{\'{\i}}nez Gonz{\'a}lez et al.(2012b)]{Marian:12_dead} Mart{\'{\i}}nez Gonz{\'a}lez, M.~J., Manso Sainz, R., Asensio Ramos, A., \& Hijano, E.\ 2012, \apj, 755, 175 

\bibitem[Mart{\'{\i}}nez Pillet et al.(2011)]{Valentin:11} Mart{\'{\i}}nez Pillet, V., Del Toro Iniesta, J.~C., {\'A}lvarez-Herrero, A., et al.\ 2011, \solphys, 268, 57 

\bibitem[Mart{\'{\i}}nez-Sykora et al.(2015)]{Sykora:15} Mart{\'{\i}}nez-Sykora, J., Moreno-Insertis, F., \& Cheung, M.~C.~M.\ 2015, \apj, 814, 2

\bibitem[Mart{\'{\i}}nez-Sykora et al.(2008)]{Sykora:08} Mart{\'{\i}}nez-Sykora, J., Hansteen, V., \& Carlsson, M.\ 2008, \apj, 679, 871

\bibitem[Massaguer \& Zahn(1980)]{Massaguer:80} Massaguer, J.~M., \& Zahn, J.-P.\ 1980, \aap, 87, 315

\bibitem[Matloch et al.(2010)]{Matloch:10} Matloch, L., Cameron, R., Shelyag, S., Schmitt, D., \& Sch\"ussler, M.\ 2010, \aap, 519, A52

\bibitem[Matloch et al.(2009)]{Matloch:09} Matloch, L., Cameron, R., Schmitt, D., \& Sch\"ussler, M.\ 2009, \aap, 504, 1041 

\bibitem[McClure et al.(2019)]{McClure:19} McClure, R.~L., Rast, M.~P., \& Mart{\'\i}nez Pillet, V.\ 2019, \solphys, 294, 18

\bibitem[Mehltretter(1978)]{Mehl:78} Mehltretter, J.~P.\ 1978, \aap, 62, 311 

\bibitem[Moreno-Insertis et al.(2018)]{Moreno:18} Moreno-Insertis, F., Martinez-Sykora, J., Hansteen, V.~H., \& Mu{\~n}oz, D.\ 2018, \apjl, 859, L26 

\bibitem[Moreno-Insertis et al.(1995)]{Moreno:95} Moreno-Insertis, F., Caligari, P. \& Sch\"ussler, M.\ 1995, \apj, 452, 894

\bibitem[M{\"u}ller et al.(2013)]{Muller:13} M{\"u}ller, D., Marsden, R.~G., St.~Cyr, O.~C., \& Gilbert, H.~R.\ 2013, \solphys, 285, 25

\bibitem[Musman(1972)]{Musman:72} Musman, S.\ 1972, \solphys, 26, 290 

\bibitem[Namba(1986)]{Namba:86} Namba, O.\ 1986, \aap, 161, 31 

\bibitem[Namba \& van Rijsbergen(1977)]{Namba:77} Namba, O., \& van Rijsbergen, R.\ 1977, Problems of Stellar Convection, 71, 119 

\bibitem[Namba \& Diemel(1969)]{Namba:69} Namba, O., \& Diemel, W.~E.\ 1969, \solphys, 7, 167

\bibitem[Nelson \& Musman(1978)]{Nelson:78} Nelson, G.~D., \& Musman, S.\ 1978, \apjl, 222, L69

\bibitem[N{\'o}brega-Siverio et al.(2016)]{Nobrega:16} N{\'o}brega-Siverio, D., Moreno-Insertis, F., \& Mart{\'{\i}}nez-Sykora, J.\ 2016, \apj, 822, 18 

\bibitem[Nordlund et al.(2009)]{Nordlund:09} Nordlund, \AA., Stein, R.~F., \& Asplund, M.\ 2009, Living Reviews in Solar Physics, 6, 2 


\bibitem[Nordlund(1985)]{Nordlund:85} Nordlund, \AA.\ 1985, \solphys, 100, 209 

\bibitem[November et al.(1981)]{November:81} November, L.~J., Toomre, J., Gebbie, K.~B., \& Simon, G.~W.\ 1981, \apjl, 245, L123 

\bibitem[Oda(1984)]{Oda:84} Oda, N.\ 1984, \solphys, 93, 243 

\bibitem[Otsuji et al.(2007)]{Otsuji:07} Otsuji, K., Shibata, K., Kitai, R., et al.\ 2007, \pasj, 59, S649


\bibitem[Orozco Su{\'a}rez et al.(2008)]{Orozco:08} Orozco Su{\'a}rez, D., Bellot Rubio, L.~R., del Toro Iniesta, J.~C., \& Tsuneta, S.\ 2008, \aap, 481, L33

\bibitem[Ortiz et al.(2014)]{Ortiz:14} Ortiz, A., Bellot Rubio, L.~R., Hansteen, V.~H., de la Cruz Rodr{\'{\i}}guez, J., \& Rouppe van der Voort, L.\ 2014, \apj, 781, 126

\bibitem[Palacios et al.(2012)]{Palacios:12} Palacios, J., Blanco Rodr{\'{\i}}guez, J., Vargas Dom{\'{\i}}nguez, S., et al.\ 2012, \aap, 537, A21 




\bibitem[Ploner et al.(2001)]{Ploner:01} Ploner, S.~R.~O., Sch{\"u}ssler, M., Solanki, S.~K., \& Gadun, A.~S.\ 2001, Advanced Solar Polarimetry -- Theory, Observation, and Instrumentation, 236, 363 

\bibitem[Ploner et al.(2000)]{Ploner:00} Ploner, S.~R.~O., Solanki, S.~K., \& Gadun, A.~S.\ 2000, \aap, 356, 1050 

\bibitem[Ploner et al.(1999)]{Ploner:99} Ploner, S.~R.~O., Solanki, S.~K., \& Gadun, A.~S.\ 1999, \aap, 352, 679 

\bibitem[Rast(2003)]{Rast:03} Rast, M.~P.\ 2003, \apj, 597, 1200 

\bibitem[Rast(1995)]{Rast:95} Rast, M.~P.\ 1995, \apj, 443, 863

\bibitem[Rempel(2018)]{Rempel:18} Rempel, M.\ 2018, \apj, 859, 161

\bibitem[Rempel(2014)]{Rempel:14} Rempel, M.\ 2014, \apj, 789, 132 

\bibitem[Requerey et al.(2017)]{Requerey:17} Requerey, I.~S., Del Toro Iniesta, J.~C., Bellot Rubio, L.~R., et al.\ 2017, \apjs, 229, 14

\bibitem[Requerey et al.(2015)]{Requerey:15} Requerey, I.~S., Del Toro Iniesta, J.~C., Bellot Rubio, L.~R., et al.\ 2015, \apj, 810, 79

\bibitem[Requerey et al.(2014)]{Requerey:14} Requerey, I.~S., Del Toro Iniesta, J.~C., Bellot Rubio, L.~R., et al.\ 2014, \apj, 789, 6

\bibitem[Riethm{\"u}ller \& Solanki(2017)]{Tino:17} Riethm{\"u}ller, T.~L., \& Solanki, S.~K.\ 2017, \aap, 598, A123 

\bibitem[Riethm{\"u}ller et al.(2014)]{Tino:14} Riethm{\"u}ller, T.~L., Solanki, S.~K., Berdyugina, S.~V., et al.\ 2014, \aap, 568, A13 

\bibitem[Rieutord et al.(2000)]{Rieutord:00} Rieutord, M., Roudier, T., Malherbe, J.~M., et al.\ 2000, \aap, 357, 1063

\bibitem[R\"osch(1960)]{Rosch:60} R\"osch, J.\ 1960, Aerodynamic Phenomena in Stellar Atmospheres, 12, 313 

\bibitem[Roth et al.(2010)]{Roth:10} Roth, M., Franz, M., Bello Gonz{\'a}lez, N., et al.\ 2010, \apjl, 723, L175 

\bibitem[Roudier et al.(2016)]{Roudier:16} Roudier, T., Malherbe, J.~M., Rieutord, M., \& Frank, Z.\ 2016, \aap, 590, A121


\bibitem[Roudier \& Muller(2004)]{Roudier:04} Roudier, T., \& Muller, R.\ 2004, \aap, 419, 757

\bibitem[Roudier et al.(2003)]{Roudier:03} Roudier, T., Ligni{\`e}res, F., Rieutord, M., Brandt, P.~N., \& Malherbe, J.~M.\ 2003, \aap, 409, 299 

\bibitem[Roudier et al.(2001)]{Roudier:01} Roudier, T., Eibe, M.~T., Malherbe, J.~M., et al.\ 2001, \aap, 368, 652


\bibitem[Ruiz Cobo \& del Toro Iniesta(1992)]{RuizIniesta:92} Ruiz Cobo, B., \& del Toro Iniesta, J.~C.\ 1992, \apj, 398, 375   

\bibitem[Simon et al.(1991)]{Simon:91} Simon, G.~W., Title, A.~M., \& Weiss, N.~O.\ 1991, \apj, 375, 775 

\bibitem[Simon \& Weiss(1991)]{SimonWeiss:91} Simon, G.~W., \& Weiss, N.~O.\ 1991, \mnras, 252, 1P 

\bibitem[Sobotka et al.(2012)]{Sobotka:12} Sobotka, M., Del Moro, D., Jur{\v{c}}{\'a}k, J., et al.\ 2012, \aap, 537, A85

\bibitem[Socas-Navarro et al.(2004)]{Hector:04} Socas-Navarro, H., Mart\'inez Pillet, V., \& Lites, B.~W.\ 2004, \apj, 611, 1139 

\bibitem[Solanki et al.(2020)]{Solanki:20} Solanki, S.~K., del Toro Iniesta, J.~C., Woch, J., et al.\ 2020, \aap, \textit{in press}

\bibitem[Solanki et al.(2017)]{Solanki:17} Solanki, S.~K., Riethm{\"u}ller, T.~L., Barthol, P., et al.\ 2017, \apjs, 229, 2 

\bibitem[Solanki et al.(2015)]{Solanki:15} Solanki, S.~K., del Toro Iniesta, J.~C., Woch, J., et al.\ 2015, Polarimetry, 305, 108 

\bibitem[Solanki et al.(2010)]{Solanki:10} Solanki, S.~K., Barthol, P., Danilovic, S., et al.\ 2010, \apjl, 723, L127 

\bibitem[Solanki et al.(1996)]{Solanki:96} Solanki, S.~K., Zufferey, D., Lin, H., Rueedi, I., \& Kuhn, J.~R.\ 1996, \aap, 310, L33 

\bibitem[Solanki(1989)]{Solanki:89} Solanki, S.~K.\ 1989, \aap, 224, 225


\bibitem[Stein \& Nordlund(1998)]{Stein:98} Stein, R.~F., \& Nordlund, \AA.\ 1998, \apj, 499, 914 

\bibitem[Stenflo(2010)]{Stenflo:10} Stenflo, J.~O.\ 2010, \aap, 517, A37 

\bibitem[Title et al.(1989)]{Title:89} Title, A.~M., Tarbell, T.~D., Topka, K.~P., et al.\ 1989, \apj, 336, 475 

\bibitem[Toriumi \& Yokoyama(2010)]{Toriumi:10} Toriumi, S., \& Yokoyama, T.\ 2010, \apj, 714, 505

\bibitem[Tortosa-Andreu \& Moreno-Insertis(2009)]{Tortosa:09} Tortosa-Andreu, A., \& Moreno-Insertis, F.\ 2009, \aap, 507, 949

\bibitem[Utz et al.(2014)]{Utz:14} Utz, D., del Toro Iniesta, J.~C., Bellot Rubio, L.~R., et al.\ 2014, \apj, 796, 79

\bibitem[van Driel-Gesztelyi \& Green(2015)]{Lidia:15} van Driel-Gesztelyi, L., \& Green, L.~M.\ 2015, Living Reviews in Solar Physics, 12, 1 

\bibitem[Vargas Dom{\'\i}nguez et al.(2010)]{Santiago:10} Vargas Dom{\'\i}nguez, S., de Vicente, A., Bonet, J.~A., et al.\ 2010, \aap, 516, A91

\bibitem[Vargas Dom{\'{\i}}nguez et al.(2014)]{Santiago:14} Vargas Dom{\'{\i}}nguez, S., Kosovichev, A., \& Yurchyshyn, V.\ 2014, \apj, 794, 140

\bibitem[V{\"o}gler \& Sch{\"u}ssler(2007)]{Vogler:07} V{\"o}gler, A., \& Sch{\"u}ssler, M.\ 2007, \aap, 465, L43

\bibitem[V{\"o}gler et al.(2005)]{Vogler:05} V{\"o}gler, A., Shelyag, S., Sch{\"u}ssler, M., et al.\ 2005, \aap, 429, 335 

\bibitem[Wang et al.(2012)]{Wang:12} Wang, J., Zhou, G., Jin, C., \& Li, H.\ 2012, \solphys, 278, 299

\bibitem[Yelles Chaouche et al.(2011)]{Lotfi:11} Yelles Chaouche, L., Moreno-Insertis, F., Mart\'inez Pillet, V., et al.\ 2011, \apjl, 727, L30 

\bibitem[Yu et al.(2011)]{Yu:11} Yu, D., Xie, Z., Hu, Q., et al.\ 2011, \apj, 743, 58

\bibitem[Zhang et al.(2009)]{Zhang:09} Zhang, J., Yang, S.-H., \& Jin, C.-L.\ 2009, Research in Astronomy and Astrophysics, 9, 921 


\end{thebibliography}
\end{document}